\documentclass[a4paper,fleqn]{cas-sc}

\usepackage[numbers,sort&compress]{natbib}

\usepackage{amsmath,amssymb,mathtools}
\usepackage{multirow}
\usepackage{booktabs}      
\usepackage{graphicx}
\usepackage{xcolor}
\usepackage{arydshln}      
\usepackage{bbding}        
\usepackage{pifont}        
\usepackage{tcolorbox}
\usepackage{float}         
\usepackage[section]{placeins} 
\usepackage{subfig}        
\usepackage{textcomp}      
\usepackage{algorithm}
\usepackage{algpseudocode}
\usepackage{listings}
\usepackage{url}

\definecolor{codegreen}{rgb}{0,0.6,0}
\definecolor{codegray}{rgb}{0.5,0.5,0.5}
\definecolor{codepurple}{rgb}{0.58,0,0.82}
\definecolor{backcolour}{rgb}{0.95,0.95,0.92}
\newfloat{listing}{tb}{lst}{}
\floatname{listing}{Listing}

\begin{document}
\let\WriteBookmarks\relax
\def\floatpagepagefraction{1}
\def\textpagefraction{.001}

\shorttitle{Metamemory Agent for Data-Free Code Generation in LLMs}
\shortauthors{S. Zhou et~al.}

\title[mode=title]{Leveraging Metamemory Agent for Enhanced Data-Free Code Generation in
Large Language Models}

\author[1]{Shengsheng Zhou}[orcid=0000-0002-5753-1043]
\ead{szhou006@e.ntu.edu.sg}
\credit{Conceptualization, Methodology, Software, Validation, Writing -- original draft}

\author[2]{Shuai Wang}[orcid=0009-0000-0481-4974]
\ead{wangshuai123@whu.edu.cn}
\credit{Methodology, Software, Investigation, Writing -- original draft}

\author[3]{Liang Ding}[orcid=0000-0001-8976-2084]
\ead{liangding.liam@gmail.com}
\credit{Methodology, Writing -- review \& editing}

\author[4]{Yibing Zhan}[orcid=0000-0003-3180-0484]
\ead{zybjy@mail.ustc.edu.cn}
\credit{Validation, Writing -- review \& editing}

\author[2]{Yong Luo}[orcid=0000-0002-2296-6370]
\cormark[1]
\ead{yluo180@gmail.com}
\credit{Conceptualization, Supervision, Funding acquisition, Writing -- review \& editing}

\author[5]{Zheng He}[orcid=0000-0002-7700-0901]
\ead{hezheng@whu.edu.cn}
\credit{Investigation, Data curation}

\author[5]{Fu Lin}[orcid=0009-0002-3749-4043]
\ead{linfu@whu.edu.cn}
\credit{Software, Visualization}

\author[6]{Dapeng Tao}[orcid=0000-0003-0783-5273]
\ead{dapeng.tao@gmail.com}
\credit{Investigation, Writing -- review \& editing}

\affiliation[1]{organization={College of Computing and Data Science, Nanyang Technological University}, country={Singapore}}
\affiliation[2]{organization={School of Computer Science, National Engineering Research Center for Multimedia Software, Wuhan University}, city={Wuhan}, country={China}}
\affiliation[3]{organization={The University of Sydney}, city={Sydney}, country={Australia}}
\affiliation[4]{organization={Yunnan United Vision Technology Company Ltd.}, country={China}}
\affiliation[5]{organization={School of Computer Science, Wuhan University}, city={Wuhan}, country={China}}
\affiliation[6]{organization={School of Information Science and Engineering, Yunnan University}, city={Kunming}, country={China}}

\cortext[1]{Corresponding author.}

\begin{abstract}
Large language models (LLMs) have shown strong performance in automated code
generation, with few-shot prompting widely used for its simplicity and
effectiveness. However, few-shot methods depend on curated or manually crafted
reference examples, limiting their applicability in data-free coding scenarios
such as real-world data-free coding scenarios and benchmarks without training
sets. Existing methods that generate reference examples via recitation or
analogy cannot guarantee their authenticity or accuracy. Inspired by human
metamemory, we propose a novel metamemory agent to enhance one-time code
generation in data-free coding scenarios. The agent guides LLMs to recall
relevant prior knowledge, evaluate confidence in recalled information, and
selectively exploit reliable content for problem solving. This agent removes the
need for external reference examples, improves the authenticity and accuracy of
recalled knowledge, and adaptively tailors the recall\&evaluation process to
each task. Extensive experiments demonstrate that the proposed metamemory agent
significantly improves one-time code generation quality across data-free coding
scenarios. The AI contribution is the metamemory agent, which makes
self-recalled examples reliable through confidence evaluation and selection; the
engineering application is data-free automated code generation, validated on
eight public benchmarks.
\end{abstract}


\maketitle

\section{Introduction}
Recently, 
large language models (LLMs) have made significant strides in understanding, generation, reasoning, and translation~\cite{fan2023automated,madani2023large,schafer2023empirical,peng2023towards,hou2024large}. By pretraining on vast amounts of textual data, the general LLMs can process language in ways that closely resemble human abilities, enabling them to perform a wide range of tasks, e.g., answering questions~\cite{shao2023prompting}, writing~\cite{van2023chatgpt}, translation~\cite{zhang2023prompting}, and summarizing documents~\cite{van2024adapted}. In many instances, general LLM's performance has matched or even exceeded human capabilities~\cite{mittelstadt2024large,strachan2024testing}.

\begin{figure}[pos=!htbp]
  \centering
  \captionsetup[subfloat]{font=footnotesize,labelfont=rm,textfont=rm}
  \subfloat[AceCoder method]
{\label{fig:acecoder_method}\includegraphics[width=0.72\textwidth]{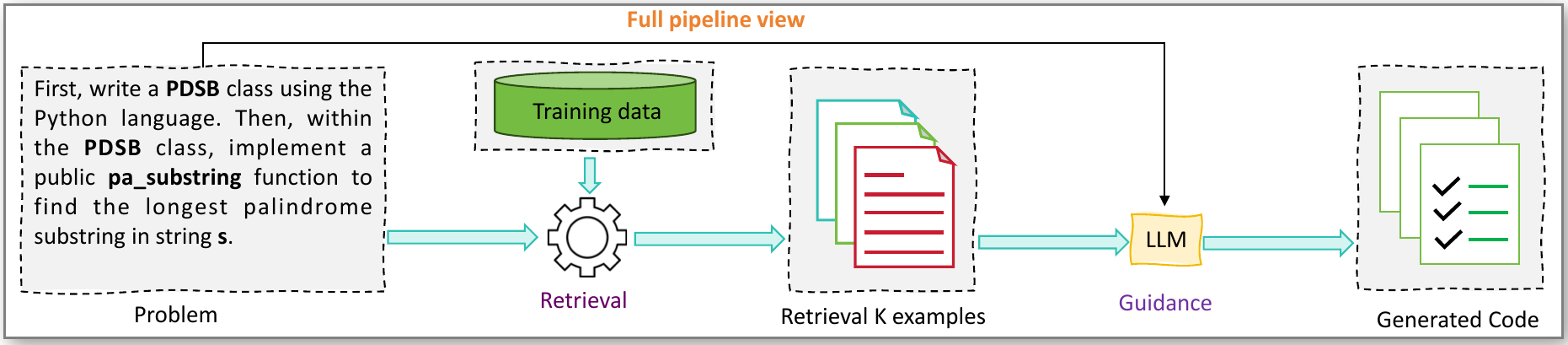}}\\
  \vspace{-3.6mm}
  \subfloat[Analogical prompting method]
{\label{fig:analogized_method}\includegraphics[width=0.72\textwidth]{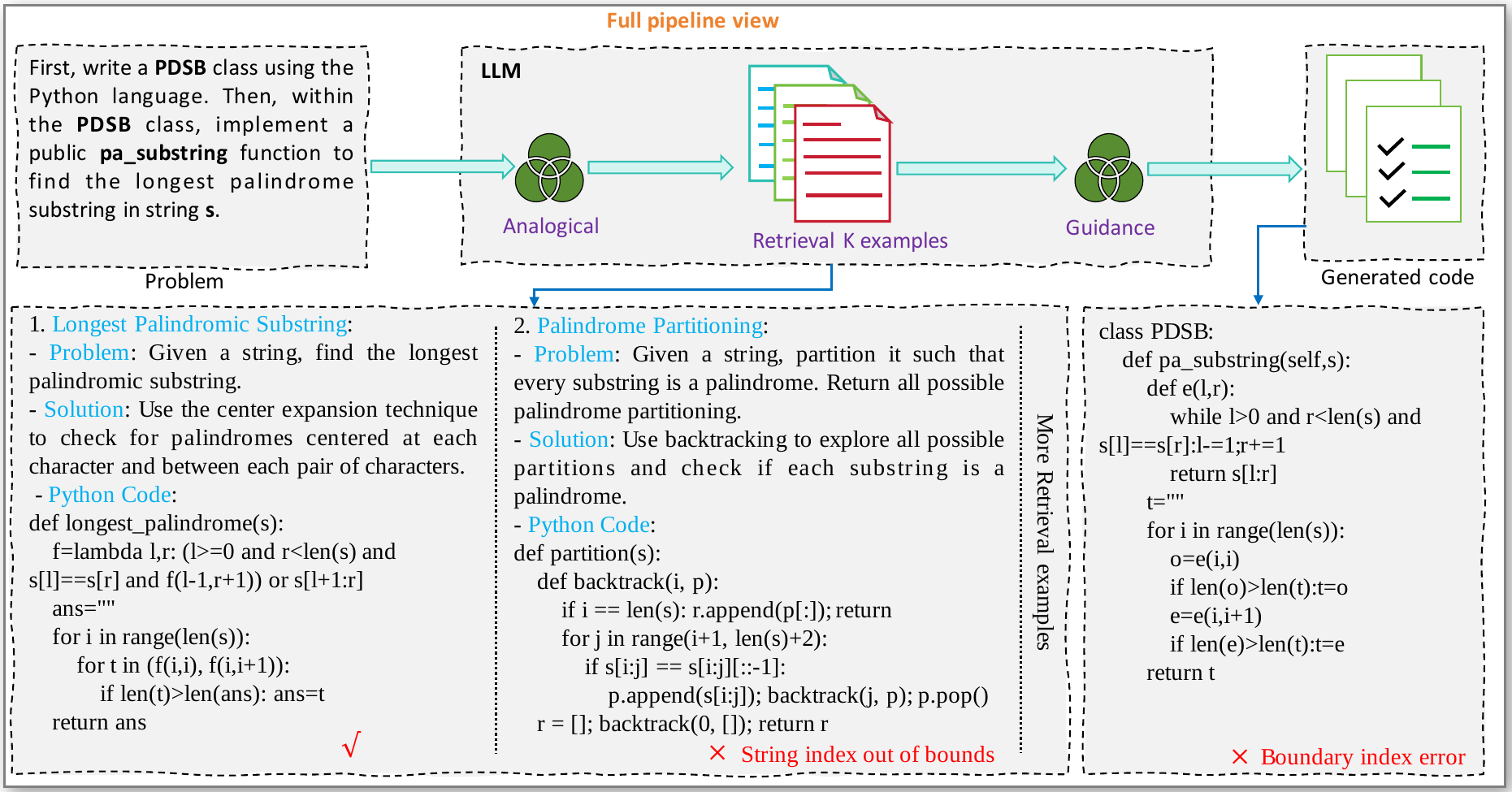}}\\
  \vspace{-3.6mm}
  \subfloat[Metamemory agent]
{\label{fig:metamemory_agent}\includegraphics[width=0.72\textwidth]{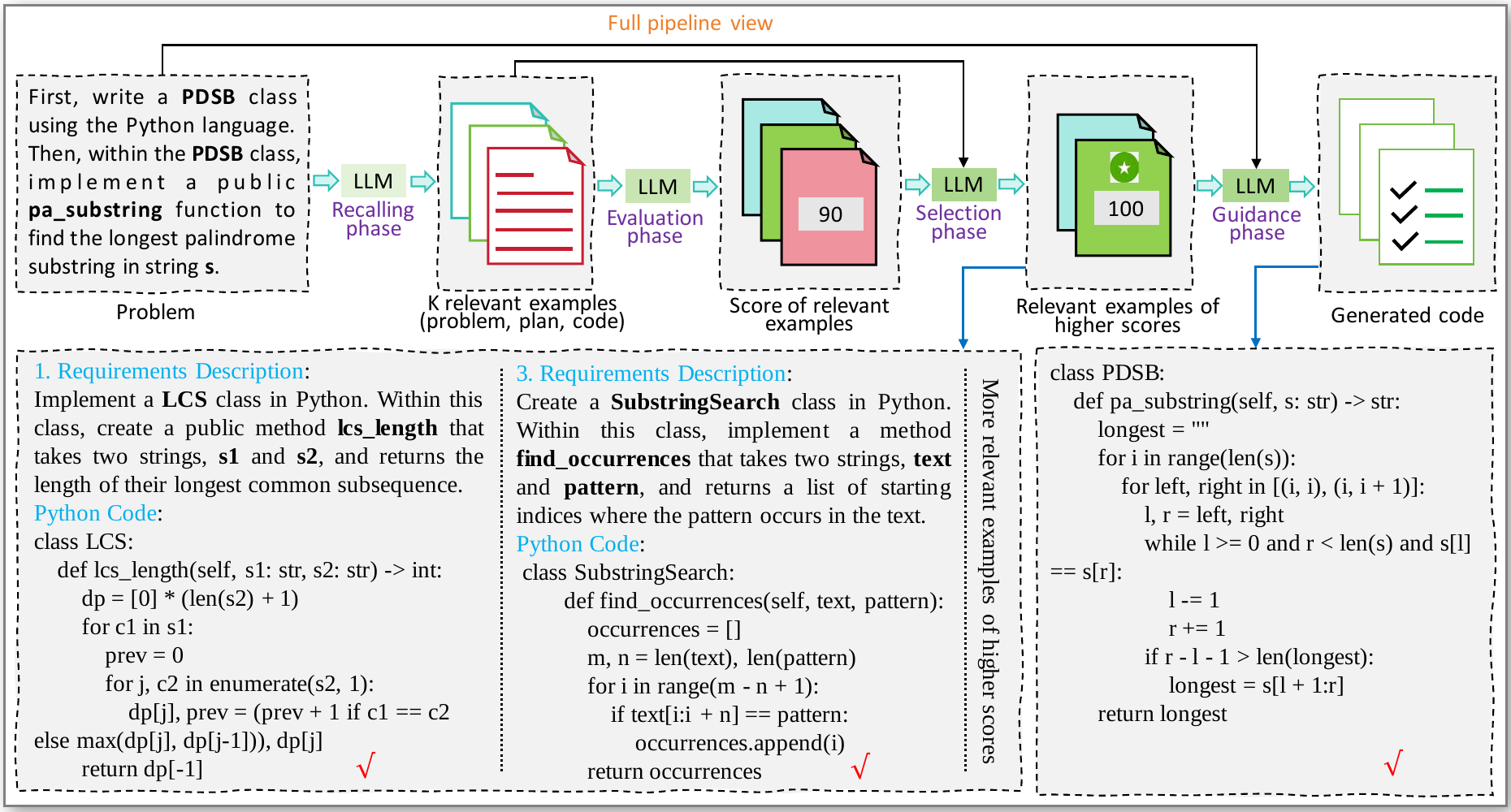}}
\vspace{-2.4mm}
  \caption{A comparison of the AceCoder method~\cite{li2023acecoder} (top), the Analogical prompting method~\cite{yasunaga2024large} (middle), and our proposed metamemory agent (bottom). We can clearly see that the AceCoder method relies on retrieving relevant examples from the training set, which poses significant challenges under real-world data-free coding scenarios with no dedicated training set. Although the analogical prompting method leverages LLMs to generate analogous examples, the accuracy of these analogous examples is often difficult to reliably guarantee. In contrast, our proposed metamemory agent not only apply real-world data-free coding scenarios with no dedicated training set but also guarantees the authenticity and accuracy of recalled relevant examples.}
  \label{fig:motive}
\end{figure}

In the field of code generation, general LLMs have demonstrated remarkable potential~\cite{li2022competition,jiang2024self,zhong2025larger,ruan2025specrover}. These general LLMs can generate code from natural language descriptions or complete and optimize existing code. This capability makes them a valuable tool for developers, particularly for tasks like automating scripts, designing algorithms, or adding code annotations~\cite{10606356,coignion2024performance,hou2023large,li2024sheetcopilot}. Since the pretraining data for the general LLMs is predominantly natural language text, with relatively little code data, they exhibit notable limitations in specialized coding scenarios~\cite{wang2024oop}. To improve performance, pretraining or fine-tuning LLMs specifically for code generation is a widely adopted approach~\cite{chen2021evaluating,roziere2023code}.
However, pretraining or fine-tuning LLMs typically requires substantial computational resources and large amounts of data, leading to high costs. Moreover, the performance of the LLMs heavily depends on the quality and diversity of the data.
Conversely, in-context learning prompting is an efficient approach that eliminates the need to pretraining or fine-tune LLMs. By carefully crafting input prompts, it guides the LLM to produce the desired output. Due to its clear advantages in terms of lower resource consumption and operational simplicity, this method is gradually replacing traditional pretraining and fine-tuning methods, emerging as an innovative solution to enhance the adaptability and efficiency of LLMs in tasks such as code generation~\cite{li2023acecoder,patel2023evaluating,geng2024large}.

Recently, few-shot prompting methods~\cite{bareiss2022code,li2023acecoder,nashid2023retrieval,yang2024improving} based on in-context learning prompting have been widely proven to significantly improve code generation tasks for LLMs (including general LLMs and code LLMs). Few-shot prompting typically adds a few reference examples to the normal prompting (i.e., the user's requirement description) to guide LLMs in addressing the requirements or problems presented in the normal prompting.
Although there are extensive researches on few-shot prompting methods~\cite{li2023acecoder,wu2023self,hongjin2022selective,peng2024revisiting}, most of them focus on retrieving the best matching reference examples from training sets to pair with the normal prompting, thereby improving the generation code performance of LLMs. Using few-shot prompting method for LLM inference is particularly difficult for real-world data-free coding scenarios or benchmark tasks (e.g., HumanEval~\cite{chen2021evaluating}, HumanEval+~\cite{liu2024your}, OOP~\cite{wang2024oop}, StudentEval~\cite{babe2023studenteval}, and LiveCodeBench~\cite{jainlivecodebench}) without training sets, as illustrated in Figure~\ref{fig:acecoder_method}.

\textit{This prompts us to consider whether we can leverage the knowledge inherent in LLMs to generate relevant reference examples, rather than relying on retrieving examples from training data or external resources to guide them in addressing the current task or problem.}
Our preliminary investigation reveals that some existing methods~\cite{sun2023recitationaugmented,yasunaga2024large} first prompt LLMs to generate relevant reference examples and then directly use these generated examples to guide the generation of LLMs. However, these methods cannot guarantee the authenticity and accuracy of the reference examples, which may consequently lead to errors in the generated code, as illustrated in Figure~\ref{fig:analogized_method}.
It is worth noting that this is especially important for programming problems. Unlike natural language, programming languages have strict syntax and semantic rules~\cite{turner1985miranda, hanford1973dynamic} that must be precisely followed for the computer to correctly understand and execute the code. \textit{So, how can we ensure the authenticity and accuracy of the reference examples}? This issue advances our research on metamemory~\cite{flavell1975metamemory,nelson1990metamemory}. Metamemory refers to an individual's awareness and regulation of their own memory processes, including the ability to recall, evaluate, and control their memory content (see Section~\ref{sec:metamemory}). 

\textit{Metamemory allows us to navigate the uncertainty of our own knowledge, transforming recall into a strategic act of problem-solving.}—— Asher Koriat~\cite{koriat1993we}

To better improve the performance of the one-time code generation of LLMs, especially for real-world data-free coding scenarios or benchmark tasks without training sets, we propose a metamemory agent inspired by the flow of human metamemory. Specifically, our proposed metamemory agent improves the performance of one-time code generation in LLMs by prompting them to use metamemory workflows.
The proposed metamemory agent consists of four main phases: recall, evaluation, selection, and guidance. Firstly, during the recall phase, the LLM is prompted to recall $K$ relevant programming problems based on the current one, along with their corresponding implementation steps (i.e., planning) and code.
Secondly, the evaluation phase prompts the LLM to assess the confidence level of each recalled programming problem and its associated code, 
Thirdly, the selection phase chooses the top $M$ examples (i.e., reference examples) with the highest confidence based on the evaluated scores. Finally, the guidance phase utilizes the top $M$ reference examples to solve the original programming problem, as illustrated in Figure~\ref{fig:metamemory_agent}. Through this metamemory agent, we can significantly enhance the authenticity and accuracy of reference examples, thereby improving the performance of the one-time code generation of LLMs.
The main contributions are summarized as follows: 
\begin{itemize}

\item Considering the flow of human metamemory, we propose a novel metamemory agent to improve one-time code generation in LLMs. This is particularly beneficial for real-world data-free coding scenarios or benchmark tasks without training sets.

\item The proposed metamemory agent has strong versatility and flexibility and can be easily plug-and-played. Moreover, the metamemory agent enhances the authenticity and accuracy of reference examples through the evaluation and screening of recalled examples and adaptively tailoring such recall\&evaluation phase process for each programming problem.

\item Extensive experimental results demonstrate that our proposed metamemory agent significantly improves the quality of LLM code generation, with the \textit{pass@$1$} score achieving an increase of 2.27\%-82.06\%. Additionally, to the best of our knowledge, this is the first attempt to integrate metamemory with LLMs.

\end{itemize}
The overall structure of this paper is organized as follows:
Section~\ref{sec:related_work} reviews related work on metamemory and code generation with LLMs; Section~\ref{sec:method} analyzes the limitations of existing in-context learning prompting methods and proposes the metamemory agent; Section~\ref{sec:experimental_setup} provides a detailed description of the experimental setup used in this study, including research questions, datasets, and LLMs; Section~\ref{sec:experimental_results} validates the effectiveness of the metamemory agent through extensive experiments; Section~\ref{sec:threats_validity} discusses the threats to validity of the metamemory agent; and Section~\ref{sec:conclusion} concludes the paper and outlines directions for future research.

\section{Related Work}
\label{sec:related_work}
\subsection{Metamemory}
\label{sec:metamemory}
Metamemory, a concept introduced by John Flavell in the 1970s, refers to an individual’s awareness and understanding of their own memory processes~\cite{flavell1975metamemory}. This cognitive ability to monitor and regulate memory emphasizes the efficient management of recalled information~\cite{lovelace1984metamemory,nelson1990metamemory,pannu2005metamemory}. It involves three key aspects: recall memory information, evaluating the accuracy of recall content, and adjusting memory strategies based on the evaluations. Through this dynamic process, individuals can improve information acquisition, enhance learning outcomes, and effectively solve complex problems~\cite{schwartz2012metamemory,miyamoto2017causal}, as illustrated in Figure~\ref{fig:metamemory}. 

\begin{center}
\includegraphics[width=0.48\textwidth]{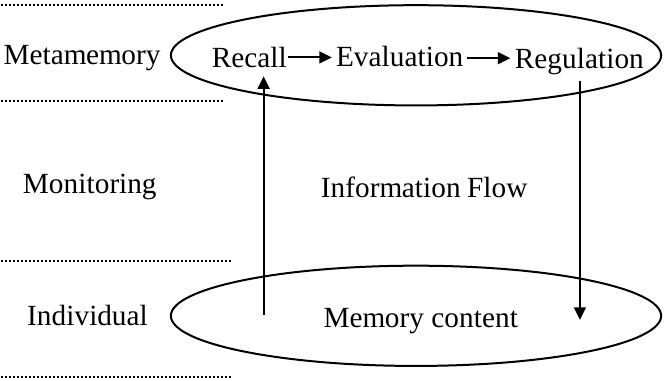}
\captionof{figure}{Metamemory workflow.}
\label{fig:metamemory}
\end{center}

Metamemory has wide applications in the field of deep learning~\cite{katzmann2019deep,wang2023mmt,li2022mm,chu2023script}. For instance, Alexander et al.~\cite{katzmann2019deep} proposed a generalized training framework based on psychological principles (namely deep metamemory), which leverages confidence estimation for efficient training enhancement and improves classification performance on unknown data. Wenjian et al.~\cite{wang2023mmt} introduced a metamemory approach to bridge the domain gap between source and target domains, effectively addressing cross-domain issues in few-shot learning. 
Jianxiang et al.~\cite{li2022mm} introduced a metamemory based few-shot object detection approach called MM-RCNN, aimed at improving object detection in optical remote sensing images. Similarly, Yang et al.~\cite{chu2023script} proposed a general framework for a metamemory driven low-resource network learning model designed to capture event correlations and predict future occurrences.

Unlike existing methods, inspired by the working mechanism of human metamemory, we propose a novel metamemory agent designed to enable LLMs to generate high-quality code in one-time through working mode of metamemory.

\subsection{Code Generation for LLMs}
\label{sec:code_generation_LLM}
The rise of LLMs has significantly advanced the development of automated code generation.
However, due to programming languages' unique syntax and semantic structures, it is particularly challenging for LLMs to generate high-quality code.
To improve the performance of automated code generation, existing research has primarily focused on pre-training or fine-tuning code LLMs~\cite{chen2021evaluating,roziere2023code,feng2020codebert,zheng2023codegeex}, post-processing methods~\cite{ding2024cycle,yu2024teaching}, and in-context learning prompting methods~\cite{li2023acecoder,wu2023self,hongjin2022selective,geng2024large}.

Pre-training and fine-tuning code LLMs~\cite{chen2021evaluating,roziere2023code,feng2020codebert,zheng2023codegeex} typically employ self-supervised learning~\cite{liu2021self,misra2020self}, leveraging large-scale code datasets (such as open-source repositories) to enhance the code LLM’s understanding by predicting the next code segment or filling in missing code. For example, in 2021, Chen et al.~\cite{chen2021evaluating} introduced the Codex model based on the GPT architecture, which was fine-tuned on publicly available GitHub~\footnote{https://github.com.} code datasets specifically for code generation tasks. Subsequently, Roziere et al.~\cite{roziere2023code} built upon the general Llama2 model~\cite{touvron2023llama} to launch the Code Llama series, designed for code generation. This series, with targeted fine-tuning, delivers high performance and supports features such as long-context processing and code completion. However, pre-training and fine-tuning code LLMs not only demand significant computational resources and time but also incur additional labor costs~\cite{min2023recent}. Moreover, the effectiveness of pre-training and fine-tuning heavily relies on the quality and diversity of the training data. If the data contains biases or noise, it may lead to the generation of inaccurate or insecure code. Excessive fine-tuning can also cause overfitting, reducing the code LLM’s generalization ability on new data. 

Post-processing (i.e., code repair) methods~\cite{ding2024cycle,yu2024teaching} primarily use secondary or multiple corrections to the code generated by LLMs. For example, Ding et al.~\cite{ding2024cycle} leveraged feedback from test cases to guide LLMs in refining initially generated code, thereby improving code quality. Yu et al.~\cite{yu2024teaching} proposed a training method based on interactive demonstrations, enabling language models to continuously optimize generated outputs through trial and error and feedback in complex tasks, achieving self-improvement. However, post-processing methods typically require LLMs to undergo multiple iterative cycles (including generating initial outputs, receiving feedback, and making revisions), significantly increasing computational costs. Additionally, these methods heavily rely on accurate and effective feedback to guide the model in achieving self-optimization.

Compared to pre-training or fine-tuning LLMs and post-processing methods, in-context learning prompting methods~\cite{li2023acecoder,wu2023self,hongjin2022selective,geng2024large} are simpler and more flexible and do not require secondary or multiple operations. Among them, few-shot prompting methods~\cite{bareiss2022code,li2023acecoder,nashid2023retrieval,yang2024improving} based on in-context learning prompting have gained widespread attention in the field of code generation due to their simple structure, flexible application, and remarkable effectiveness, e.g., AceCoder~\cite{li2023acecoder}. However, few-shot prompting methods rely on retrieving a few reference examples from a training set or manually crafting reference shots, as illustrated in Figure~\ref{fig:acecoder_method}. 

Unlike existing few-shot prompting methods, our proposed metamemory agent doesn't require retrieving or manually crafting reference examples. Instead, it generates relevant reference examples through the LLM's recall phase and enhances their accuracy via the evaluation phase, as illustrated in Figure~\ref{fig:metamemory_agent}.

\section{Method}
\label{sec:method}
\subsection{overview}
In this part, we first define some symbols. Then, we formalize several related in-context learning prompting methods (i.e., normal prompting, few-shot prompting~\cite{bareiss2022code,li2023acecoder,nashid2023retrieval,yang2024improving}, analogical prompting~\cite{yasunaga2024large}) and analyze their drawbacks. 

\noindent $\bullet$ \textbf{Symbol definition}.
We denote the reasoning layer of the pre-trained LLM with parameters $\theta$ as $\textit{f}$, and lowercase letters \textbf{\textit{x}}, \textbf{\textit{y}}, $\cdots$ to denote a language sequence, i.e., $\textbf{\textit{x}}=(\textit{x}[1],\cdots,\textit{x}[n])$, where each \textit{x}[i] is a token. Note: in the formalization process, we omit the representation of the encoding and decoding processes, focusing solely on the representation of the reasoning process in LLMs.

\noindent $\bullet$ \textbf{Normal prompting (NP)}.
The normal prompting is using the LLM to map the input $\mathbf{\textit{x}}$ (i.e., the user's requirements) to $\mathbf{\textit{y}}$. This process can be formalized as: $\textbf{\textit{y}} \sim \textit{f}^{NP}_{\theta}(\textbf{\textit{y}}|\textbf{\textit{x}})$. Although LLMs can directly meet user requirements by inputting $\textbf{\textit{x}}$, their understanding is far less profound than that of humans. Especially, in complex coding scenarios or when clear prompts are lacking, the LLM's output may remain superficial and fail to address the core of the programming problem~\cite{wang2023chatcoder}.

\noindent $\bullet$ \textbf{Few-shot prompting (FSP)}. Few-shot prompting uses a few examples retrieved from the training set (or manually created examples) as references to guide an LLM in generating responses that meet the user's requirements. This process can be formalized as: $\textbf{\textit{y}} \sim \textit{f}^{FSP}_{\theta}(\textbf{\textit{y}}|\textbf{\textit{c}},\textbf{\textit{x}})$, where $\textbf{\textit{c}}$ represents a small number of reference examples. Although few-shot prompting is highly flexible and effective~\cite{bareiss2022code,li2023acecoder,nashid2023retrieval,yang2024improving}, it is especially challenging to use for LLM inference in real-world data-free coding scenarios or benchmark tasks without training sets, as illustrated in Figure~\ref{fig:acecoder_method}.

\noindent $\bullet$ \textbf{Analogical prompting (AP)}. The analogical prompting uses manually designed instructions $\textbf{\textit{d}}$ to help the LLM identify relevant examples $\textbf{\textit{c}}_1, \cdots, \textbf{\textit{c}}_n$ to the user's requirements. Then, the analogous examples $\textbf{\textit{c}}_1, \cdots, \textbf{\textit{c}}_n$ (i.e., reference examples) are used to guide the LLM in generating results that meet the user's requirements. This process can be formalized as: $\textbf{\textit{y}} \sim \textit{f}^{AP}_{\theta}(\textbf{\textit{y}}|\textbf{\textit{d}},\textbf{\textit{c}}_{1},\cdots,\textbf{\textit{c}}_{n}\textbf{\textit{x}})$. Although analogical prompting can generate relevant examples to the user's requirements, it cannot guarantee the authenticity and accuracy of the analogized reference examples, i.e., $\textbf{\textit{c}}_1, \cdots, \textbf{\textit{c}}_n$, as illustrated in Figure~\ref{fig:analogized_method}.

To improve the quality of one-time code generation in real-world data-free scenarios or benchmark tasks without training sets, we propose a novel metamemory agent inspired by the metamemory mechanism in Section~\ref{sec:metamemory}. Our proposed metamemory agent mainly consists of four phases: recall, evaluation, selection, and guidance, with its overall framework demonstrated in Figure~\ref{fig:metamemory_agent}.


\begin{lstlisting}[language=Python, escapeinside={(*@}{@*)}, caption= {Prompt template for recalling phase.}, label={lst:recall_agent}]
Your task is to recall {K} related problems based on the given original problem. 
These related problems should be thematically related but not identical to the 
original, sharing similar algorithmic patterns, data structure requirements, or 
problem-solving approaches.
(*@\textbf{\# Original Problem:}@*)
{question}
(*@\textbf{\# Requirements for Each Recalled Problem:}@*)
1. (*@<description>@*): A clear, concise problem statement that highlights its relevance to the original problem;
2. (*@<code>@*): Complete, executable code with no missing components;
3. (*@<planning>@*): Step-by-step solution approach;
(*@\textbf{\# Output Format (XML):}@*)
Your response must strictly follow this format with {recall_K} problem entries:
(*@\textbf{<root>}@*)
  (*@\textbf{<problem>}@*)
  (*@\textbf{<description>}@*)
  # Problem description here.
  (*@\textbf{</description>}@*)
  (*@\textbf{<code>}@*)
  # Complete {language} code here.
  (*@\textbf{</code>}@*)
  (*@\textbf{<planning>}@*)
  # Planning to solve this problem.
  (*@\textbf{</planning>}@*)
  (*@\textbf{</problem>}@*)
  <!-- Repeat <problem> block for remaining {K - 1} problems -->
(*@\textbf{</root>}@*)
\end{lstlisting}

\subsection{Metamemory agent}
\subsubsection{Recall}
Recalling is a cognitive process through which individuals retrieve previously acquired information from memory. This process typically involves two core stages: retrieval and extraction.
During retrieval, individuals access stored information in response to new stimuli, and during extraction, they selectively utilize relevant portions of the retrieved content to support reasoning or decision-making. Inspired by this mechanism, the recalling stage adopts a similar strategy, wherein \textit{recalling relevant examples} serves as the initial phase, mirroring how humans access related experiences when faced with novel problems. This recall phase not only facilitates the LLM’s comprehension of the retrieved content but also provides contextualized analogical guidance for solving new tasks. Specifically, given a new programming problem $\textbf{\textit{x}}$, the recall phase prompts the LLM to retrieve $K$ semantically or structurally similar problems from its memory. Furthermore, our recall phase expands on the principle of the analogical method~\cite{yasunaga2024large} by guiding the LLM to generate planning steps for these recall-related programming problems, e.g.,
\begin{equation}
\label{eq:recall_problem}
\left\{\hat{\textbf{\textit{x}}}^i, \hat{\textbf{\textit{y}}}^i, \hat{\textbf{\textit{s}}}^i\right\}_{i=1}^K 
\sim f^{re}_\theta\Big(\textbf{\textit{x}} \;\Big|\; \textbf{\textit{d}}_{re}, history \Big),
\end{equation}
where \textit{re} is an abbreviation for \textit{recalling}; $\hat{\textbf{\textit{x}}}^i$ represents the relevant programming problem of the $i$-th recollection; $\hat{\textbf{\textit{s}}}^i$ represents the planning for the $i$-th related programming problem; $\hat{\textbf{\textit{y}}}^i$ represents the code for the $i$-th related programming problem; \textit{history} represents the previously recalled information, including the earlier generated $\left\{\hat{\textbf{\textit{x}}}^{i-1}, \hat{\textbf{\textit{y}}}^{i-1}, \hat{\textbf{\textit{s}}}^{i-1}\right\}$; $\textbf{\textit{d}}_{re}$ represents the instruction requirements for recalling related programming problems. More specifically, our instruction prompt template $\textbf{\textit{d}}_{re}$ for recalling related examples is shown in Listing~\ref{lst:recall_agent}.

\begin{lstlisting}[language=Python, escapeinside={(*@}{@*)}, caption= {Prompt template for evaluation phase.}, label={lst:evaluation_agent}]
Given a competitive programming problem and a {language} code solution, carefully 
evaluate its correctness using the following criteria:
1. Does the code correctly implement the required algorithm?
2. Does it handle all edge cases and constraints specified in the problem?
3. Is the code free of syntax errors and logical flaws?
4. Does it produce the correct output for typical and edge-case inputs?

(*@\textbf{\# Recalled Problem:}@*)
{Description of the recall problem}
(*@\textbf{\# Code:}@*)
{code}
(*@\textbf{\# Planning:}@*)
{planning}

(*@\textbf{\# Evaluation Guidelines:}@*)
- In <explanation>, provide a detailed assessment including:
  - Identification of any errors (syntax, logic, edge case handling);
  - Analysis of whether the planning approach matches the problem requirements;
- In <confidence>, provide an integer score (0-100) representing your confidence 
in the planning(*@\textquotesingle@*)s correctness, where:
  - 0-30: Severe issues, planning is fundamentally incorrect;
  - 31-60: Partial correctness but significant flaws;
  - 61-90: Mostly correct with minor issues;
  - 91-100: Correct and robust implementation;

--------------------------------
Your response must strictly follow this xml format:
(*@\textbf{<root>}@*)
(*@\textbf{<explanation>}@*) 
# Detailed assessment here.
(*@\textbf{</explanation>}@*)     
(*@\textbf{<confidence>}@*)
# 0-100 integer score here. 
(*@\textbf{</confidence>}@*)
(*@\textbf{</root>}@*)
\end{lstlisting}

\subsubsection{Evaluation}
After extracting recalled information, humans typically engage in a self-evaluation process to evaluate the accuracy of their memory, a cognitive function often referred to as metamemory. This evaluation involves estimating the accuracy and relevance of the retrieved content, guided by prior experience and memory heuristics. In alignment with this cognitive process, the metamemory agent incorporates a confidence evaluation stage, wherein the LLM is instructed to assess the accuracy of each recalled programming example $\left\{\hat{\textbf{\textit{x}}}^i,
\hat{\textbf{\textit{y}}}^i,
\hat{\textbf{\textit{s}}}^i\right\}$, including the description, code, and planning of relevant programming requirements. To implement this, we design prompts that generate confidence scores between 0 and 100, e.g.,
\begin{equation}
\label{eq:evaluation_problem}
\textbf{\textit{C}}^i \sim f^{ev}_\theta\Big(\hat{\textbf{\textit{x}}}^i, \hat{\textbf{\textit{y}}}^i, \hat{\textbf{\textit{s}}}^i \;\big|\; \textbf{\textit{d}}_{ev}\Big), 
\; i=1,2,\dots,K, \; \textbf{\textit{C}}^i \in [0,100],
\end{equation}
where \textit{ev} is an abbreviation for \textit{evaluation}; $\textbf{\textit{C}}^i$ represents the confidence score for the $i$-th related programming problem; $\textbf{\textit{d}}_{ev}$ represent the evaluation instruction requirements. Our evaluation instruction prompt template $\textbf{\textit{d}}_{ev}$ is shown in Listing~\ref{lst:evaluation_agent}.

\begin{lstlisting}[language=Python, escapeinside={(*@}{@*)}, caption= {Prompt template for selection phase.}, label={lst:selection_agent}]
Here are {K} examples, each containing a requirement description, a code, planning
, and the corresponding confidence score. Please select the {M} examples with the 
highest scores and respond only with their numbers, such as ``Example 1'', 
``Example 2'', without adding any explanation.

(*@\textbf{\# Example 1:}@*)
## Requirement Description
{Description of the recall problem}
## Code
{code}
## Planning
{planning}
## Confidence Score
{score}
# More Recall Examples...
\end{lstlisting}

\subsubsection{Selection}
After completing the evaluation phase, humans typically filter retrieved memories to determine which ones are most relevant and reliable for the current reasoning task. This process involves not only assessing the accuracy and completeness of the memory content but also evaluating its relevance to the target problem (i.e., the original problem). Inspired by this cognitive mechanism, we design a selection phase within the metamemory agent. This phase primarily selects the most valuable reference examples from the set of recalled items based on confidence scores $\textbf{\textit{C}}^i$ obtained during the prior evaluation. By prioritizing high-confidence examples, the phase enables LLMs to focus on information that is most likely to provide correct guidance, thereby improving the accuracy and robustness of their outputs. At the same time, the selection phase effectively suppresses or excludes recalled examples that may contain logical flaws, semantic errors, or low relevance to the current task, providing high-quality references for reasoning on new problems. This significantly enhances the decision-making ability and reliability of LLMs in complex tasks. The selection phase can be formalized as:
\begin{equation}
\label{eq:select_problem}
\left\{\tilde{\textbf{\textit{x}}}^j, \tilde{\textbf{\textit{y}}}^j, \tilde{\textbf{\textit{s}}}^j\right\}_{j=1}^M 
\sim f^{se}_\theta\Big(\{\hat{\textbf{\textit{x}}}^i, \hat{\textbf{\textit{y}}}^i, \hat{\textbf{\textit{s}}}^i, \textbf{\textit{C}}^i\}_{i=1}^K \;\big|\; \textbf{\textit{d}}_{se}\Big), 
\; M \le K,
\end{equation}
where \textit{se} is an abbreviation for \textit{selection}; $M$ represents the number of reference examples selected; $\textbf{\textit{d}}_{se}$ represent the selected instruction requirements. The instruction prompt template $\textbf{\textit{d}}_{se}$ for the selection phase is shown in Listing~\ref{lst:selection_agent}.


\subsubsection{Guidance}
The guidance phase represents the final component of the metamemory agent and serves as a critical step in solving the target programming problem (i.e., the original programming problem). During this phase, the selected high-confidence examples $\left\{\tilde{\textbf{\textit{x}}}^j, \tilde{\textbf{\textit{y}}}^j, \tilde{\textbf{\textit{s}}}^j\right\}$, including both the code and the corresponding solution plans, provide the LLM with structured and reliable guidance to generate the final code solution for the target problem $\textbf{\textit{x}}$, i.e.,
\begin{equation}
\label{eq:last_result}
\textbf{\textit{y}} \sim f^{gu}_\theta\Big(\{\tilde{\textbf{\textit{x}}}^j, \tilde{\textbf{\textit{y}}}^j, \tilde{\textbf{\textit{s}}}^j\}_{j=1}^M \;\big|\; \textbf{\textit{d}}_{gu}\Big),
\end{equation}
where \textit{gu} is an abbreviation for \textit{guidance}; $\textbf{\textit{d}}_{gu}$ represents the instruction requirements for writing the code for the original programming problem. The instruction prompt template $\textbf{\textit{d}}_{gu}$ is shown in Listing~\ref{lst:guidance_agent}.

\begin{lstlisting}[language=Python, escapeinside={(*@}{@*)}, caption= {Prompt template for guidance phase.}, label={lst:guidance_agent}]
(*@\textbf{\# Example 1:}@*)
## Input:
{Related programming problem}
## Response:
{code}
{planning}
# More Examples...

(*@\textbf{\# New Input:}@*)
## Input:
{problem}
## Response:
\end{lstlisting}





\section{Experimental Setup}
\label{sec:experimental_setup}
\subsection{Research Questions}
In this section, we will explore the effectiveness of the metamemory agent and compare it with Normal prompting, CoT prompting, Plan-and-Solve prompting, Analogical prompting, and Few-shot prompting methods. Our experiments aim to answer the following research questions:

\noindent $\bullet$ \textbf{RQ1. Overall Performance}: How does the LLM based on our proposed metamemory agent perform in real-world data-free and data-available coding scenarios? This RQ1 aims to investigate whether the metamemory agent can improve performance in real-world data-free and data-available coding scenarios.

\noindent $\bullet$ \textbf{RQ2. Performance on Open-source and Closed-source LLMs}: How do the open-source and closed-source LLMs based on our proposed metamemory agent perform in real-world data-free and data-available coding scenarios? This RQ2 aims to investigate whether both open-source and closed-source LLMs based on the metamemory agent can achieve performance improvements in real-world data-free and data-available coding scenarios.

\noindent $\bullet$ \textbf{RQ3. Performance of LLMs with different parameters}: How do LLMs with larger and smaller numbers of parameters based on the metamemory agent perform in real-world data-free and data-available coding scenarios? This RQ3 aims to investigate whether both larger parameter and smaller parameter LLMs based on the metamemory agent can achieve performance improvements in real-world data-free and data-available coding scenarios.

\noindent $\bullet$ \textbf{RQ4. Metamemory Agent VS. Few-shot prompting}: How does the proposed metamemory agent perform in comparison to few-shot prompting methods (e.g., AceCoder~\cite{li2023acecoder})? This RQ4 aims to investigate whether the metamemory agent achieves better improvement than the few-shot prompting methods.

\noindent $\bullet$ \textbf{RQ5. Authenticity and Accuracy of the Reference Example}: Does our proposed metamemory agent ensure the authenticity and accuracy of the reference examples? This RQ5 aims to investigate whether the metamemory agent can improve the authenticity and accuracy of reference examples.

\noindent $\bullet$ \textbf{RQ6. Impact of Recalled Examples}: What impact does increasing the number of recalled related examples have on the performance of the metamemory agent? This RQ6 aims to investigate whether recalling more examples continuously improves the performance of LLMs in code generation tasks.

\noindent $\bullet$ \textbf{RQ7. Impact of Selected Examples}: What impact does selecting multiple recalled examples with higher confidence have on the performance of our proposed metamemory agent? This RQ7 aims to investigate whether selecting more recalled examples can enhance the performance of LLMs in code generation tasks.

\noindent $\bullet$ \textbf{RQ8. Performance in other programming languages}: Can our proposed metamemory agent be applied to multilingual benchmarks? This RQ8 aims to investigate whether the metamemory agent can improve performance on code generation tasks in other programming languages (e.g., C++, C\#, Java, etc).

\noindent $\bullet$ \textbf{RQ9. Case study}: 
Why does our proposed metamemory agent can enhance the performance of LLMs in one-time code generation, especially in real-world data-free coding scenarios. This RQ9 aims to investigate the reasons why the metamemory agent improves the one-time code generation performance of LLMs.

\subsection{Datasets}
To better evaluate the performance of the metamemory agent, we selected eight benchmarks, namely HumanEval~\cite{chen2021evaluating}, HumanEval+~\cite{liu2024your}, OOP~\cite{wang2024oop}, StudentEval~\cite{babe2023studenteval}, MBPP~\cite{austin2021program}, LiveCodeBench~\cite{jainlivecodebench}, codeforces~\cite{li2022competition,yasunaga2024large}, MultiPL-E~\cite{cassano2023multipl}.

\noindent
$\bullet$ \textbf{HumanEval~\cite{chen2021evaluating}}: The HumanEval benchmark, proposed by OpenAI, is an important benchmark for evaluating the code generation capabilities of LLMs. It consists of 164 Python-based programming tasks, each including a function definition, a natural language description, and corresponding unit tests. The benchmark is designed to assess whether a model can generate syntactically correct and functionally accurate executable code based on natural language instructions. The HumanEval benchmark employs metrics such as \textit{pass@$k$} to measure the probability of producing correct solutions across multiple generations, making it one of the core standards for evaluating the performance of code generation models.

\noindent
$\bullet$ \textbf{HumanEval+~\cite{liu2024your}}: The HumanEval+ benchmark is an extended and enhanced version of OpenAI’s original HumanEval~\cite{chen2021evaluating} benchmark, designed to offer a more rigorous and comprehensive evaluation environment for code generation models. Based on the original 164 programming tasks, it expands and refines the test cases to cover a broader range of edge cases and potential error scenarios, enabling a more accurate assessment of the robustness and generalizability of the generated code.

\noindent
$\bullet$ \textbf{OOP~\cite{wang2024oop}}:
The OOP benchmark is a code generation benchmark specifically designed for LLMs, aimed at evaluating their performance on object-oriented programming tasks. It includes 431 Python programming problems covering key object-oriented programming concepts such as class definitions, method implementation, attribute encapsulation, inheritance, and polymorphism.

\noindent
$\bullet$ \textbf{StudentEval~\cite{babe2023studenteval}}: The StudentEval benchmark, released in 2023, is a test benchmark containing 1,675 prompts across 48 questions, written by 80 students who had completed only one semester of a Python programming course.
The StudentEval benchmark identifies four key disjoint subsets for each question participant: first success (the correct code was generated on the first attempt), first failure (the first attempt was unsuccessful, and the participant moved on to the next problem), final success (the correct code was generated on the final attempt), and final failure (the final attempt was unsuccessful, and the participant moved on to the next problem).

\noindent
$\bullet$ \textbf{MBPP~\cite{austin2021program}}: The MBPP dataset is a benchmark collection of Python programming problems, primarily aimed at beginner programmers, covering fundamental concepts such as string manipulation, list operations, loops, and conditional statements. It uses natural language prompts to evaluate program synthesis and code generation tasks, provides complete test cases for automatic verification, and is widely used to assess the coding capabilities of LLMs.

\noindent
$\bullet$ \textbf{LiveCodeBench~\cite{jainlivecodebench}}: LiveCodeBench is a comprehensive and contamination-free benchmark platform specifically designed to evaluate the code generation capabilities of LLMs. By continuously collecting newly released problems from programming competitions such as LeetCode, it ensures that the evaluation remains unaffected by existing training data, thereby providing a more authentic and dynamic assessment of code generation performance. In the experiments, we chose to use the latest version, namely ``release\_v6''.

\noindent
$\bullet$ \textbf{Codeforces~\cite{yasunaga2024large}}: The Codeforces benchmark is a dataset collected by Michihiro et al.,~\cite{yasunaga2024large} based on prior work~\cite{kulal2019spoc,li2022competition}, from the \textit{\url{codeforces.com}} website. It covers multiple difficulty levels, including $A,B,\cdots$. In the experiment, we followed Yasunaga et al.~\cite{yasunaga2024large} and used level-A problems as the test set for evaluation.

\noindent
$\bullet$ \textbf{MultiPL-E~\cite{babe2023studenteval}}:
The MultiPL-E is a multilingual code generation benchmark designed to evaluate the ability of LLMs to generate code across different programming languages. The benchmark is based on the HumanEval~\cite{chen2021evaluating} and MBPP~\cite{austin2021program} Python task benchmarks and translates these tasks into 18 programming languages, including Java, C++, JavaScript, Rust, Go, Swift, and others.

\subsection{Baselines}
To demonstrate that the metamemory agent can significantly improve the performance of LLMs in generating code on one-time, especially for benchmark tasks (e.g., HumanEval, HumanEval+, OOP, StudentEval, LiveCodeBench, and MultiPL-E) without training sets, we have compared the metamemory agent with several baselines and state-of-the-art methods.

\noindent
$\bullet$ \textbf{Normal prompting} (i.e., the user's requirements) instructs the LLM to generate the corresponding code directly without adding any extra instructions.

\noindent
$\bullet$ \textbf{CoT prompting}~\cite{kojima2022large} adds ``Let's think step by step'' after the normal prompting, instructing the LLM to generate the final code step by step.

\noindent
$\bullet$ \textbf{Plan-and-Solve prompting}~\cite{wang-etal-2023-plan} first formulates an overall plan that decomposes the complex task into several more manageable subtasks. Then, the LLM executes these subtasks sequentially according to the plan, thereby improving the quality of the final generated output.

\noindent
$\bullet$ \textbf{Analogical prompting}~\cite{yasunaga2024large} instructs LLMs to provide multiple relevant examples based on the original question, and then use these relevant examples to solve the original problem.

\subsection{Evaluation Metric}
To objectively and fairly evaluate the performance of the metamemory agent, we adopt appropriate evaluation metrics across different benchmarks. For the HumanEval, HumanEval+, StudentEval, LiveCodeBench, Codeforces, and MultiPL-E benchmarks, we follow Chen et al.~\cite{chen2021evaluating} and Shinn et al.~\cite{shinn2024reflexion} and use the \textit{pass@$k$} metric for evaluation. For the OOP benchmark, we follow Wang et al.~\cite{wang2024oop} and use both the \textit{pass@$k$} and \textit{pass@$o$} metrics. For the Codeforces benchmark, we follow Li et al.~\cite{li2022competition} and adopt the \textit{n@$k$} metric. The calculation of \textit{pass@$k$} is as follows:
\begin{equation}
\label{eq:mertric_pass@k}
\textit{pass@$k$}:=\mathop{\mathbb{E}}_{Problems} \left[1-\frac{{\binom{n-c}{k}}}{\binom{n}{k}} \right].
\end{equation}
In Eq. (\ref{eq:mertric_pass@k}), $n$ represents the number of code generations for a given problem; $c$ represents the number of generated $n$ codes that passed the test.

The calculation formula for the \textit{pass@$o$} metric is:
\begin{align}
\label{eq:total_add_pass_matching}
\alpha = \sum_{i=1}^{n} f\left(X_i\right), where f(X_i)=\left\{
\begin{aligned}
    1,& \, if \, utf\left(X_i\right) \, passed \, and \, \sum_{j}^{m} x_j \exists X_i \\
    0,& \, \mathrm{otherwise}
\end{aligned}
\right.,
\end{align}

\begin{equation}
\label{eq:mertric_pass@o}
\textit{pass@$o$}:=\mathop{\mathbb{E}}_{Problems} \left[1-\frac{{\binom{n-\alpha}{k}}}{\binom{n}{k}} \right].
\end{equation}
In Eq. (\ref{eq:total_add_pass_matching}), $X_i$ represents the $i$-${th}$ generated code samples; $\alpha$ represents the number of $n$ generated code samples that pass the unit test and achieve successful matching; $ut\left(\cdot\right)$ represents the unit test function; $m$ denotes the number of key points in the current natural language description; and $x_j$ denotes the $j$-${th}$ keyword points in natural language.

\subsection{Models}
To better validate the performance of the metamemory agent, we evaluate four open-source LLMs, i.e., Qwen2.5-7B-Instruct~\cite{qwen25}, Qwen2.5-Coder-7B-Instruct~\cite{hui2024qwen2}, InternLM2.5-7B-Chat~\cite{cai2024internlm2} and Gemma-3-4B-IT~\cite{team2025gemma}, and two closed-source LLMs, i.e., ChatGPT~\cite{achiam2023gpt} and GPT-4o mini~\footnote{~\url{https://platform.openai.com/docs/models/gpt-4o-mini}}.

\noindent
$\bullet$ \textbf{Qwen2.5-7B-Instruct~\cite{qwen25}}:
Qwen2.5-7B-Instruct is the latest instruction-tuned version of Alibaba’s Qwen series of LLMs. It contains approximately 7.61 billion parameters and supports a context window of up to 128K tokens. The LLM is optimized for complex instruction following, long-text generation, and multitasking, demonstrating outstanding performance in language understanding, multilingual capabilities, programming, mathematical reasoning, and logical inference.

\noindent
$\bullet$ \textbf{Qwen2.5-Coder-7B-Instruct~\cite{hui2024qwen2}}: Qwen2.5-Coder-7B-Instruct is an instruction-tuned model in Alibaba’s Qwen series, specifically designed for programming tasks. It contains approximately 7.61 billion parameters and supports a context window of up to 128K tokens. The LLM covers 92 programming languages and is specially optimized for code generation, auto-completion, error correction, code reasoning, and multilingual programming tasks.

\noindent
$\bullet$ \textbf{InternLM2.5-7B-Chat~\cite{cai2024internlm2}}: InternLM2.5-7B-Chat is the core conversational model of the open-source LLM series InternLM2.5, developed by the Shanghai AI Lab. With 7 billion parameters, the model is designed for practical use, showing great ability in reasoning, understanding long texts, and using tools. It supports a context length of up to 32,768 tokens, enabling efficient handling of complex conversations, multi-turn interactions, and tool-calling tasks.

\noindent
$\bullet$ \textbf{Gemma-3-4B-IT~\cite{team2025gemma}}: Gemma-3-4B-IT is the instruction-tuned version of Google’s open-source LLM series Gemma 3, featuring 4 billion parameters. Built upon Gemini 2.0 technology, it is designed for efficient deployment and optimization across multimodal tasks. The model excels at text generation, image understanding, question answering, information summarization, and reasoning, and it can manage very long conversations and interactions between text and images thanks to its ability to process up to 128,000 tokens at once.

\noindent
$\bullet$ \textbf{ChatGPT~\cite{achiam2023gpt}}: ChatGPT is a language model developed by OpenAI based on a generative pre-trained transformer model. ChatGPT can understand and generate natural language text, supporting a variety of applications, including dialogue generation, text composition, and creative writing.

\noindent
$\bullet$ \textbf{GPT-4o mini}: GPT-4o mini is a compact and efficient multimodal LLM released by OpenAI in July 2024, serving as a streamlined version of the GPT-4o series. The model is optimized for cost efficiency and inference speed, offering exceptional capabilities in text generation, image understanding, speech processing, and complex reasoning. It supports a context window of up to 128K tokens and an output length of up to 16K tokens.

The detailed description of the LLMs for Qwen2.5-7B-Instruct, Qwen2.5-Coder-7B-Instruct, InternLM2.5-7B-Chat, Gemma-3-4B-IT, ChatGPT and GPT-4o mini is shown in Table~\ref{tab:code_models}.

\begin{table}[htbp]
  \centering
  \caption{Overview of the evaluated models.}
  \resizebox{0.99\linewidth}{!}{
    \begin{tabular}{cccccc}
    \toprule
    Model & Organization & Time  & Open-source & Size & Source \\
    \midrule
    Qwen2.5-7B-Instruct & Qwen Team & 2024  & \Checkmark & 7B & \url{https://huggingface.co/Qwen/Qwen2.5-7B-Instruct} \\
    Qwen2.5-Coder-7B-Instruct & Qwen Team & 2024  & \Checkmark & 7B & \url{https://huggingface.co/Qwen/Qwen2.5-Coder-7B-Instruct} \\
    InternLM2.5-7B-Chat & Shanghai AI Lab & 2024  & \Checkmark & 7B & \url{https://huggingface.co/internlm/internlm2_5-7b-chat} \\
    Gemma-3-4B-IT & Google DeepMind & 2025 & \Checkmark & 4B &\url{https://huggingface.co/google/gemma-3-4b-it} \\
    ChatGPT & OpenAI & 2021  & \ding{55} & 175B & \url{https://platform.openai.com/docs/models/gpt-3-5-turbo} \\
    GPT-4o mini & OpenAI  & 2024 & \ding{55} & - & \url{https://platform.openai.com/docs/models/gpt-4o-mini} \\
    \bottomrule
    \end{tabular}%
    }
  \label{tab:code_models}%
\end{table}%

\subsection{Parameter Settings}
In the experimental setup, the temperature of the LLMs was set to 0.1 and the top-p value to 0.95 for evaluations on the HumanEval, HumanEval+, MBPP, OOP, StudentEval, Codeforces, and MultiPL-E benchmarks. For the LiveCodeBench benchmark, the temperature was adjusted to 0.2 while keeping top-p at 0.95. The open-source LLMs used were all evaluated in an environment equipped with four NVIDIA A100 GPUs.

\section{Experimental Results}
\label{sec:experimental_results}
\subsection{Results and Analysis}
To evaluate the performance of the metamemory agent in real-world data-free coding scenarios or benchmark tasks without training sets, we conducted experiments on the HumanEval, HumanEval+, StudentEval, LiveCodeBench, and OOP benchmarks, with the results presented in Tables~\ref{tab:result01} and~\ref{tab:result_studenteval}, respectively. At the same time, to verify that the metamemory agent is also effective for benchmarks with training data, we included the results on the MBPP benchmark in Table~\ref{tab:result01}. Additionally, we used the Qwen2.5-Coder-7B-Instruct and GPT-4o mini as references and compared the metamemory agent with few-shot prompting methods on the Codeforces benchmark, with the results shown in Table~\ref{tab:codeforces}.

\subsubsection{RQ1. Overall Performance}
From Tables~\ref{tab:result01} and~\ref{tab:result_studenteval}, it is clear that the metamemory agent significantly outperforms baseline methods across five data-free coding scenarios and one data-available coding scenario.
Taking Qwen2.5-7B-Instruct as an example, in the data-free coding tasks, its performance on the HumanEval, HumanEval+, and LiveCodeBench benchmarks reaching 85.49, 78.00, and 14.81, which correspond to improvements of 4.63\%, 8.41\%, and 6.32\% over Normal prompting.
On the OOP benchmark, the metamemory agent also surpasses Normal prompting in both \textit{pass@$k$} and \textit{pass@$o$} metrics (achieving 41.07 and 18.10, respectively).
Meanwhile, on the StudentEval benchmark, its average performance exceeds that of Normal prompting by 4.97\%.
In addition, in the five data-free coding scenarios, Qwen2.5-7B-Instruct with the metamemory agent consistently outperforms CoT prompting, Plan-and-Solve prompting, and Analogical prompting. In the data-available coding scenario, Qwen2.5-7B-Instruct with the metamemory agent achieves a performance score of 76.78, which is also higher than Normal prompting (72.29), CoT prompting (75.31), Plan-and-Solve prompting (64.83), and Analogical prompting (60.57).
Similarly, the performance of other LLMs (i.e., Qwen2.5-Coder-7B-Instruct, InternLM2.5-7B-Chat, Gemma-3-4B-IT, ChatGPT, and GPT-4o mini) based on the metamemory agent also exceed those of the baseline methods. These results indicate that the metamemory agent can not only significantly enhance the performance of LLMs in data-free coding scenarios but also improve their effectiveness in data-driven coding scenarios.

\noindent
\textbf{Answer to RQ1:} In five data-free coding scenarios and one data-available coding scenario, LLMs (i.e., Qwen2.5-7B-Instruct, Qwen2.5-Coder-7B-Instruct, InternLM2.5-7B-Chat, Gemma-3-4B-IT, ChatGPT, and GPT-4o mini) integrated with our proposed metamemory agent consistently outperform the baseline methods (i.e., Normal prompting, CoT prompting, Plan-and-Solve prompting, Analogical prompting).

\begin{table}[htbp]
  \centering
  \caption{The performance of LLMs (e.g., Qwen2.5-7B-Instruct, etc) using different methods on the HumanEval, HumanEval+, MBPP, LiveCodeBench, and OOP benchmarks, where HumanEval, HumanEval+, LiveCodeBench, and OOP are benchmarks without training data. On the OOP benchmark, the left side shows the \textit{pass@$k$} results, while the right side shows the \textit{pass@$o$} results. The Bold text indicates the best result. The red color indicates the relative improvement of our metamemory agent over Normal prompting. In the experiment, we set $K=3$ and $M=1$.}
  \resizebox{0.99\linewidth}{!}{
    \begin{tabular}{cccccccc}
    \toprule
    LLMs  & Methods & HumanEval & HumanEval+ & MBPP  & LiveCodeBench & \multicolumn{2}{c}{OOP} \\
    \midrule
    \multirow{5}[2]{*}{Qwen2.5-7B-Instruct} & Normal prompting & 81.71     & 71.95     & 72.29     & 13.93     & 41.07 & 18.10 \\
          & CoT prompting & 82.93     & 75.61     & 75.31     & 14.53     & 40.14     & 16.47 \\
          & Plan-and-Solve prompting & 81.71     & 71.95     & 64.83     & 13.75     & 33.41     & 11.37 \\
          & Analogical prompting & 73.78     & 67.07     & 60.57     & 12.97     & 38.98     & 13.69 \\
          & Metamemory agent & \textbf{85.49}     & \textbf{78.00}     & \textbf{76.78}     & \textbf{14.81}    & \textbf{42.82}      & \textbf{19.69} \\
    \midrule
    \multicolumn{2}{c}{Improvement} & \textcolor{red}{\textbf{4.63\%}}     & \textcolor{red}{\textbf{8.41\%}}     & \textcolor{red}{\textbf{6.21\%}}     & \textcolor{red}{\textbf{6.32\%}}     & \textcolor{red}{\textbf{4.26\%}}     & \textcolor{red}{\textbf{8.78\%}} \\
    \midrule
    \multirow{5}[2]{*}{Qwen2.5-Coder-7B-Instruct} & Normal prompting & 85.37     & 79.88     & 64.48     & 14.31     & 46.17     & 18.10 \\
          & CoT prompting & 87.02     & 81.10     & 63.98     & 14.44     & 44.08     & 15.55 \\
          & Plan-and-Solve prompting & 86.59     & 76.83     & 54.16    & 13.81     & 46.38    & 16.47 \\
          & Analogical prompting & 50.00     & 45.73     & 28.46     & 13.43     & 31.09     & 11.60 \\
          & Metamemory agent & \textbf{88.41}     & \textbf{83.49}     & \textbf{73.80}     & \textbf{14.88}     & \textbf{47.68}     & \textbf{20.15} \\
    \midrule
    \multicolumn{2}{c}{Improvement} & \textcolor{red}{\textbf{3.56\%}}     & \textcolor{red}{\textbf{4.52\%}}     & \textcolor{red}{\textbf{14.45\%}}     & \textcolor{red}{\textbf{3.98\%}}     & \textcolor{red}{\textbf{3.27\%}}     & \textcolor{red}{\textbf{11.33\%}} \\
    \midrule
    \multirow{5}[2]{*}{InternLM2.5-7B-Chat} & Normal prompting & 60.98     & 54.88     & 40.05     & 5.50     & 17.67     & 7.42 \\
          & CoT prompting & 61.59     & 57.32     & 46.93     & 6.51     & 19.95     & 7.84 \\
          & Plan-and-Solve prompting & 73.17     & 64.63     & 30.98     & 6.12     & 17.63     & 8.58 \\
          & Analogical prompting & 72.56     & 64.02     & 36.02     & 5.64     & 18.56     & 7.42 \\
          & Metamemory agent & \textbf{76.07}     & \textbf{66.20}     & \textbf{48.34}     & \textbf{6.82}     & \textbf{20.56}     & \textbf{9.12} \\
    \midrule
    \multicolumn{2}{c}{Improvement} & \textcolor{red}{\textbf{24.75\%}}     & \textcolor{red}{\textbf{20.63\%}}     & \textcolor{red}{\textbf{20.70\%}}     & \textcolor{red}{\textbf{24.00\%}}     & \textcolor{red}{\textbf{16.36\%}}     & \textcolor{red}{\textbf{22.91\%}} \\
    \midrule
    \multirow{5}[2]{*}{Gemma-3-4B-IT} & Normal prompting & 68.00     & 60.98     & 72.29     & 8.53     & 18.56     & 9.05 \\
          & CoT prompting & 70.73     & 62.98     & 73.29     & 8.80     & 32.71     & 12.53 \\
          & Plan-and-Solve prompting & 69.51     & 60.98     & 67.51     & 8.64     & 29.23     & 14.15 \\
          & Analogical prompting & 69.51     & 61.58     & 70.71     & 8.48     & 22.27     & 6.73 \\
          & Metamemory agent & \textbf{71.12}     & \textbf{63.59}     & \textbf{74.52}     & \textbf{8.87}     & \textbf{33.79}     & \textbf{15.78} \\
    \midrule
    \multicolumn{2}{c}{Improvement} & \textcolor{red}{\textbf{4.59\%}}     & \textcolor{red}{\textbf{4.28\%}}     & \textcolor{red}{\textbf{3.08\%}}     & \textcolor{red}{\textbf{3.99\%}}     & \textcolor{red}{\textbf{82.06\%}}     & \textcolor{red}{\textbf{74.36\%}} \\
    \midrule
    \multirow{5}[2]{*}{ChatGPT} & Normal prompting   & 86.59     & 78.10     & 79.60     & 15.63   & 47.80     & 23.67\\
          & CoT prompting & 89.02     & 83.54     & 81.82     & 16.08     & 56.61     & 35.03 \\
          & Plan-and-Solve prompting & 89.63     & 81.10     & 68.51     & 16.27     & 57.31     & 32.02 \\
          & Analogical prompting & 84.15     & 77.44     & 77.83     & 15.80     & 58.47     & 29.70 \\
          & metamemory agent & \textbf{90.20}     & \textbf{84.27}     & \textbf{82.29}     & \textbf{16.45}     & \textbf{59.44}     & \textbf{37.61} \\
    \midrule
    \multicolumn{2}{c}{Improvement} & \textcolor{red}{\textbf{4.17\%}}     & \textcolor{red}{\textbf{7.90\%}}     & \textcolor{red}{\textbf{3.38\%}}     & \textcolor{red}{\textbf{5.25\%}}     & \textcolor{red}{\textbf{23.72\%}}     & \textcolor{red}{\textbf{58.89\%}} \\
    \midrule
    \multirow{5}[2]{*}{GPT-4o mini} & Normal prompting & 86.58     & 81.09     & 79.34     & 15.64     & 53.13     & 34.57 \\
          & CoT prompting & 88.41     & 82.93     & 80.82     & 15.96     & 56.38     & 35.50 \\
          & Plan-and-Solve prompting & 88.41     & 81.71     & 69.02     & 15.70     & 59.40     & 34.57 \\
          & Analogical prompting & 85.98     & 78.64     & 76.32     & 15.60     & 59.63     & 30.39 \\
          & Metamemory agent & \textbf{89.20}     & \textbf{82.93}     & \textbf{82.06}     & \textbf{16.40}     & \textbf{60.06}     & \textbf{37.47} \\
    \midrule
    \multicolumn{2}{c}{Improvement} & \textcolor{red}{\textbf{3.03\%}}     & \textcolor{red}{\textbf{2.27\%}}     & \textcolor{red}{\textbf{3.43\%}}     & \textcolor{red}{\textbf{4.86\%}}     & \textcolor{red}{\textbf{3.04\%}}     & \textcolor{red}{\textbf{8.39\%}} \\
    \bottomrule
    \end{tabular}%
    }
  \label{tab:result01}%
\end{table}%

\begin{table}[htbp]
  \centering
  \caption{The performance of LLMs (e.g., Qwen2.5-7B-Instruct, etc)  using different methods on the StudentEval benchmark. Bold text indicates the best result. The $\boldsymbol{\Delta}\left(\uparrow\right)$ indicates the relative improvement of our metamemory agent over normal prompting, CoT prompting, Plan-and-Solve prompting, and Analogical prompting. In the experiment, we set $K=3$ and $M=1$.}
  \resizebox{0.99\linewidth}{!}{
    \begin{tabular}{cccccccc}
    \toprule
    Model & Method & First Failure & First Success & Last Failure & Last Success & Avg & $\boldsymbol{\Delta}\left(\uparrow\right)$ \\
    \midrule
    \multicolumn{1}{c}{\multirow{5}[0]{*}{Qwen2.5-7B-Instruct}} & Normal prompting & 18.86	& 75.94	& 21.35	& 77.84  & 48.50  & \textcolor{red}{\textbf{4.97\%}}  \\
    & CoT prompting &18.61	&83.42	&20.79	&72.43  & 48.81  & \textcolor{red}{\textbf{4.30\%}}  \\
    & Plan-and-Solve prompting & 18.61	& 80.75	& 20.79	& 72.97  & 48.28  & \textcolor{red}{\textbf{5.45\%}}  \\
    & Analogical prompting & 18.36	& 75.40	& 16.85	& 65.95  & 44.14  & \textcolor{red}{\textbf{15.34\%}}  \\
    & Metamemory agent & \textbf{20.35}	& \textbf{83.87}	& \textbf{20.82}	& \textbf{78.58}  & \textbf{50.91}  & -  \\
    \hdashline 
    \multicolumn{1}{c}{\multirow{5}[0]{*}{Qwen2.5-Coder-7B-Instruct}} & Normal prompting & 21.84	& 85.56	& 23.60	& 78.92  & 52.48  & \textcolor{red}{\textbf{1.52\%}}  \\
    & CoT prompting & 22.58	& 83.42	& 22.47	& 74.05  & 50.63  & \textcolor{red}{\textbf{5.23\%}}  \\
    & Plan-and-Solve prompting & 18.11	& 74.87	& 20.22	& 75.68  & 47.21  & \textcolor{red}{\textbf{12.86\%}}  \\
    & Analogical prompting & 12.66	& 42.25	& 12.36	& 45.41  & 28.17  & \textcolor{red}{\textbf{89.14\%}}  \\
    & Metamemory agent & \textbf{23.09}  & \textbf{85.61}  & \textbf{24.91}  & \textbf{79.51}  & \textbf{53.28}  & -  \\
    \hdashline 
    \multicolumn{1}{c}{\multirow{5}[0]{*}{InternLM2.5-7B-Chat}} & Normal prompting & 15.38	& 72.19	& 13.48	& 64.86  & 41.48  & \textcolor{red}{\textbf{10.05\%}}  \\
    & CoT prompting & 16.13	& 75.94	& 16.29	& 67.57  & 43.98  & \textcolor{red}{\textbf{3.80\%}}  \\
    & Plan-and-Solve prompting & 15.63	& 74.87	& 16.85	& 69.73  & 44.28  & \textcolor{red}{\textbf{3.09\%}}  \\
    & Analogical prompting & 12.66	& 64.17	& 14.04	& 50.81  & 35.42  & \textcolor{red}{\textbf{28.88\%}}  \\
    & Metamemory agent & \textbf{16.14}	& \textbf{77.80}	& \textbf{17.73}	& \textbf{70.92}  & \textbf{45.65}  & -  \\
    \hdashline 
    \multicolumn{1}{c}{\multirow{5}[0]{*}{Gemma-3-4B-IT}} & Normal prompting & 17.62 & 68.98 & 17.98 & 62.16  & 41.44  & \textcolor{red}{\textbf{2.90\%}}  \\
    & CoT prompting & 16.63	& 67.38	& 14.61	& 60.54  & 39.79  & \textcolor{red}{\textbf{7.16\%}}  \\
    & Plan-and-Solve prompting & 16.13	& 62.57	& 15.73	& 56.22  & 37.66  & \textcolor{red}{\textbf{13.22\%}}  \\
    & Analogical prompting & 16.38	& 64.71	& 13.48	& 58.38  & 38.24  & \textcolor{red}{\textbf{11.51\%}}  \\
    & Metamemory agent & \textbf{18.11}	& \textbf{69.45}	& \textbf{18.61}	& \textbf{64.38}  & \textbf{42.64}  & -  \\
    \hdashline 
    \multicolumn{1}{c}{\multirow{5}[0]{*}{ChatGPT}} & Normal prompting & 22.33	& 81.28	& 25.28	& 82.70  & 52.90  & \textcolor{red}{\textbf{8.41\%}}  \\
    & CoT prompting & 21.59	& 81.82	& 26.97	& 81.62  & 53.00  & \textcolor{red}{\textbf{8.21\%}}  \\
    & Plan-and-Solve prompting & 23.08	& 82.89	& 27.53	& 82.16  & 53.92  & \textcolor{red}{\textbf{6.36\%}}  \\
    & Analogical prompting & 21.84	& 75.40	& 25.28	& 77.30  & 49.96  & \textcolor{red}{\textbf{14.79\%}}  \\
    & Metamemory agent & \textbf{32.51}	& \textbf{84.61}	& \textbf{27.97}	& \textbf{84.30}  & \textbf{57.35}  & -  \\
    \hdashline 
    \multicolumn{1}{c}{\multirow{5}[0]{*}{GPT-4o mini}} & Normal prompting & 24.81 & 81.82	& 25.84	& 83.24  & 53.93  & \textcolor{red}{\textbf{3.54\%}}  \\
    & CoT prompting & 22.58	& 81.82	& 27.53	& 82.70  & 53.66  & \textcolor{red}{\textbf{4.06\%}}  \\
    & Plan-and-Solve prompting & 25.58	& 82.35	& 25.84	& 82.16  & 53.98  & \textcolor{red}{\textbf{3.45\%}}  \\
    & Analogical prompting & 20.60	& 79.14	& 26.97	& 77.84  & 51.14  & \textcolor{red}{\textbf{9.19\%}}  \\
    & Metamemory agent & \textbf{25.83}	& \textbf{85.03}	& \textbf{28.28}	& \textbf{84.22}  & \textbf{55.84}  & -  \\
    \bottomrule
    \end{tabular}%
    }
  \label{tab:result_studenteval}%
\end{table}%

\subsubsection{RQ2. Performance on Open-source and Closed-source LLMs}
As shown in Tables~\ref{tab:result01} and~\ref{tab:result_studenteval}, the InternLM2.5-7B-Chat (open-source LLM) with the metamemory agent achieves performance scores of 76.07, 66.20, 48.34, and 6.82 on HumanEval, HumanEval+, MBPP, and LiveCodeBench, respectively. Compared to the InternLM2.5-7B-Chat using Normal prompts, this represents improvements of 24.75\%, 20.63\%, 20.70\%, and 24.00\%. The model exhibits similar performance gains on the OOP and StudentEval benchmarks.
Similarly, the GPT-4o mini (closed-source LLM)) with the metamemory agent achieves scores of 89.20, 82.93, 82.06, and 16.40 on HumanEval, HumanEval+, MBPP, and LiveCodeBench, respectively, representing improvements of 3.03\%, 2.27\%, 3.43\%, and 4.86\% compared to using Normal prompting. This model also shows similar improvements on the OOP and StudentEval benchmarks.
These results demonstrate that the metamemory agent can significantly enhance the performance of both open-source and closed-source LLMs.

\noindent
\textbf{Answer to RQ2:} Our proposed metamemory agent can effectively enhance the performance of both open-source and closed-source LLMs in data-free and data-available coding scenarios.

\subsubsection{RQ3. Performance of LLMs with different parameters}
As shown in Tables~\ref{tab:result01} and~\ref{tab:result_studenteval}, the large-parameter ChatGPT with the metamemory agent achieves scores of 90.20, 82.93, 82.06, 16.40, and 57.35 on the HumanEval, HumanEval+, MBPP, LiveCodeBench, and StudentEval benchmarks, respectively. Compared to ChatGPT using Normal prompting, this represents performance improvements of 4.17\%, 7.90\%, 3.38\%, 5.25\%, and 8.41\%. On the OOP benchmark, the LLM attains 59.44 on \textit{pass@$k$} and 37.61 on \textit{pass@$o$}, corresponding to increases of 23.72\% and 58.89\% over the Normal prompting. For the smaller-parameter Gemma-3-4B-IT, employing the metamemory agent leads to improvements of 4.59\%, 4.28\%, 3.08\%, 3.99\%, and 2.90\% on HumanEval, HumanEval+, MBPP, LiveCodeBench, and StudentEval, with similarly strong results on the OOP benchmark. These findings indicate that the metamemory agent can substantially enhance the performance of both large-parameter and small-parameter LLMs in data-free and data-available coding scenarios.

\noindent
\textbf{Answer to RQ3:} Our proposed metamemory agent can effectively enhance the performance of both large-parameter and small-parameter LLMs in data-free and data-available coding scenarios.

\subsubsection{RQ4. Metamemory Agent VS. Few-shot prompting}
As shown in Table~\ref{tab:codeforces}, the Qwen2.5-Coder-7B-Instruct with the metamemory agent achieves a score of 29.49 on the Codeforces benchmark, which corresponds to improvements of 27.33\%, 27.11\%, and 24.54\% over 1-shot CoT prompting, AceCoder (1-shot, Level B), and AceCoder (1-shot, Level B \& C), respectively. On GPT-4o mini, the metamemory agent achieves 29.96, while 1-shot CoT prompting, AceCoder (1-shot, Level B), and AceCoder (1-shot, Level B \& C) score 23.82, 23.67, and 24.12, showing improvements of about 25.78\%, 26.57\%, and 24.21\%. These results indicate that, compared to few-shot prompting, the metamemory agent can more effectively enhance the performance of LLMs on code generation tasks.

\noindent
\textbf{Answer to RQ4:} Compared to few-shot prompting, our proposed metamemory agent can more effectively enhance the performance of LLMs on code generation tasks.

\begin{table}[htbp]
  \centering
  \caption{Performance of the Qwen2.5-Coder-7B-Instruct and GPT-4o mini with different prompting methods (e.g., AceCoder) on the Codeforces benchmark. Bold text shows the best result. Level B and Level C are different levels of data in the Codeforces benchmark. During the experiment, we used Level B and Level C as the AceCoder retrieval data and set $K=3$ and $M=1$ for the metamemory agent.}
  \resizebox{0.99\linewidth}{!}{
    \begin{tabular}{ccccc|ccccc}
    \toprule
    \multirow{2}[4]{*}{LLM} & \multirow{2}[4]{*}{Methods} & \multicolumn{2}{c}{\textit{Pass@$k$}} & \multirow{2}[4]{*}{Avg} & \multirow{2}[4]{*}{LLM} & \multirow{2}[4]{*}{Methods} & \multicolumn{2}{c}{\textit{Pass@$k$}} & \multirow{2}[4]{*}{Avg} \\
\cmidrule{3-4}\cmidrule{8-9}          &       & $k=1$   & $k=10$   &       &       &       & $k=1$   & $k=10$   &  \\
    \midrule
    \multirow{11}[2]{*}{Qwen2.5-Coder-7B-Instruct} & Normal prompting & 13.95     & 29.91     & 21.93     & \multirow{11}[2]{*}{GPT-4o mini} & Normal prompting & 15.12     & 30.35     & 22.74 \\
          & CoT prompting & 14.27     & 31.03     & 22.65     &       & CoT prompting & 15.43     & 31.77     & 23.60 \\
          & 1-shot CoT prompting & 14.86     & 31.45     & 23.16     &       & 1-shot CoT prompting & 15.62     & 32.01     & 23.82 \\
          & Analogical prompting & 16.32     & 33.92     & 25.12     &       & Analogical prompting & 17.00     & 34.27     & 25.64 \\
          & AceCoder (1-shot, Level B) & 15.02     & 31.37     & 23.20     &       & AceCoder (1-shot, Level B) & 15.50     & 31.83     & 23.67 \\
          & AceCoder (2-shot, Level B) & 16.00     & 33.88     & 24.94     &       & AceCoder (2-shot, Level B) & 16.97     & 34.47     & 25.72 \\
          & AceCoder (3-shot, Level B) & 17.21     & 34.69     & 25.95     &       & AceCoder (3-shot, Level B) & 17.71     & 35.86     & 26.79 \\
          & AceCoder (1-shot, Level B \& C) & 15.34     & 32.01     & 23.68     &       & AceCoder (1-shot, Level B \& C) & 15.83     & 32.40     & 24.12 \\
          & AceCoder (2-shot, Level B \& C) & 16.87     & 35.23     & 26.05     &       & AceCoder (2-shot, Level B \& C) & 17.31     & 36.06     & 26.69 \\
          & AceCoder (3-shot, Level B \& C) & 18.14     & 39.03     & 28.59     &       & AceCoder (3-shot, Level B \& C) & 18.88     & 39.75     & 29.32 \\
          & Metamemory agent & \textbf{18.83}     & \textbf{40.15}     & \textbf{29.49}     &       & Metamemory agent & \textbf{19.00}     & \textbf{40.91}     & \textbf{29.96} \\
    \bottomrule
    \end{tabular}%
    }
  \label{tab:codeforces}%
\end{table}%

\begin{figure}[htbp]
  \centering
  \captionsetup[subfloat]{font=footnotesize,labelfont=rm,textfont=rm}
  \subfloat[HumanEval]
  {\label{fig:stdy1}\includegraphics[scale=0.32]{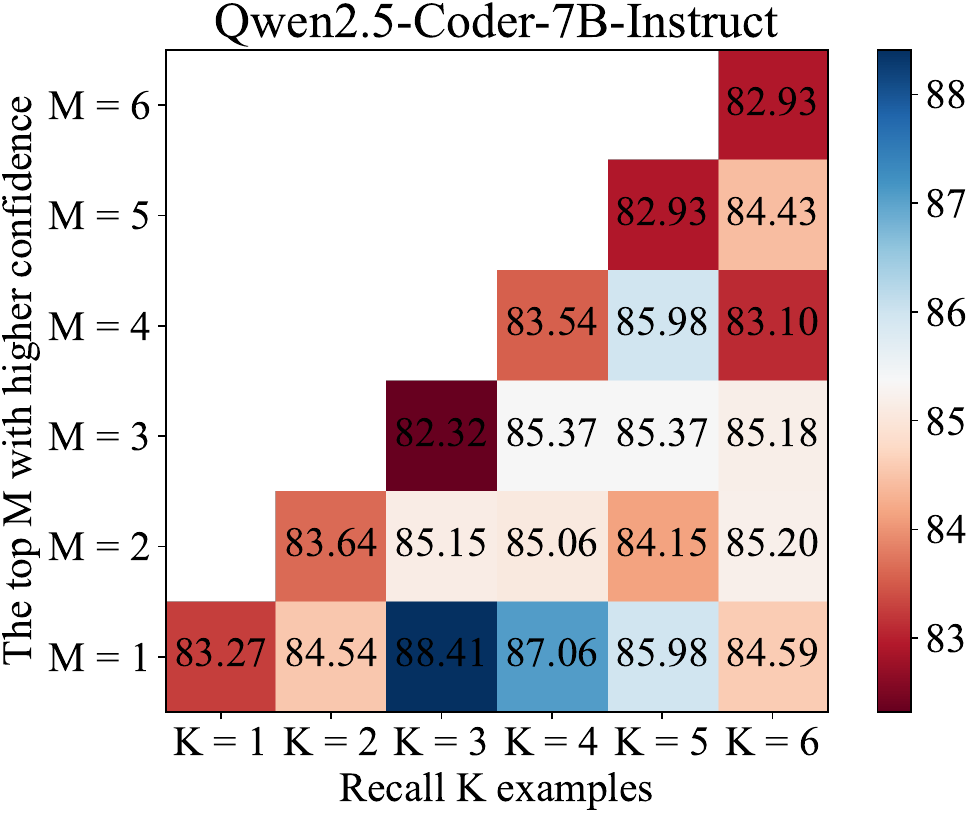}}
  \hspace{2.0mm}
  \subfloat[HumanEval]
  {\label{fig:stdy2}\includegraphics[scale=0.32]{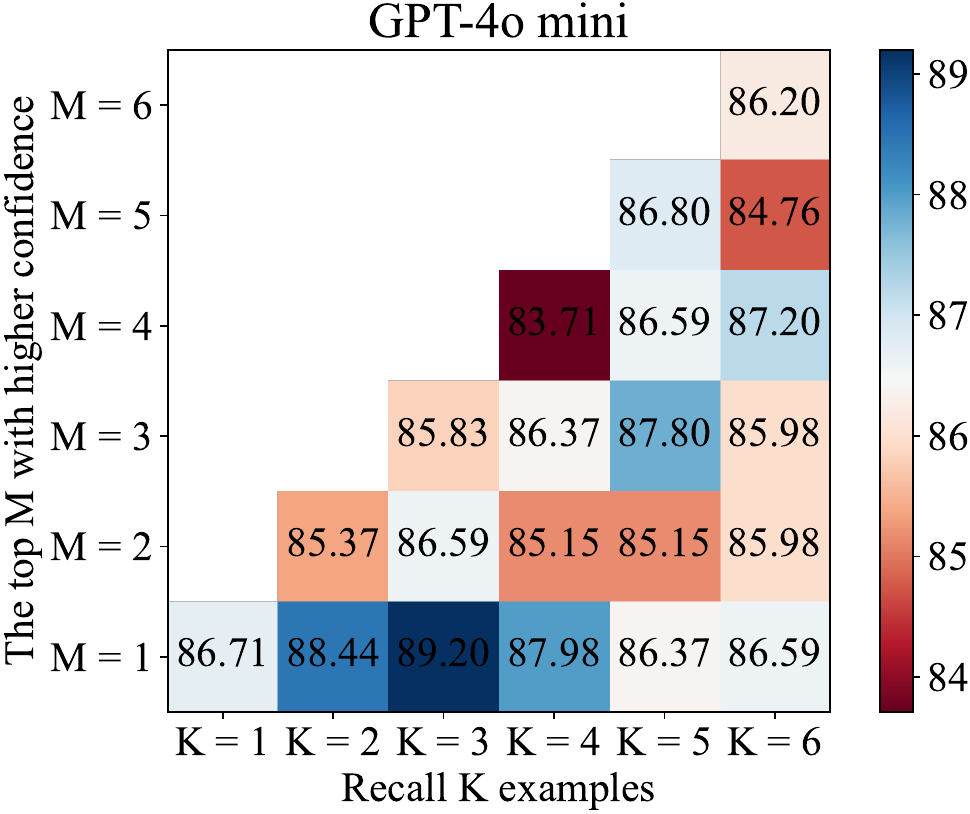}}
  \\
 \subfloat[OOP]
  {\label{fig:stdy3}\includegraphics[scale=0.32]{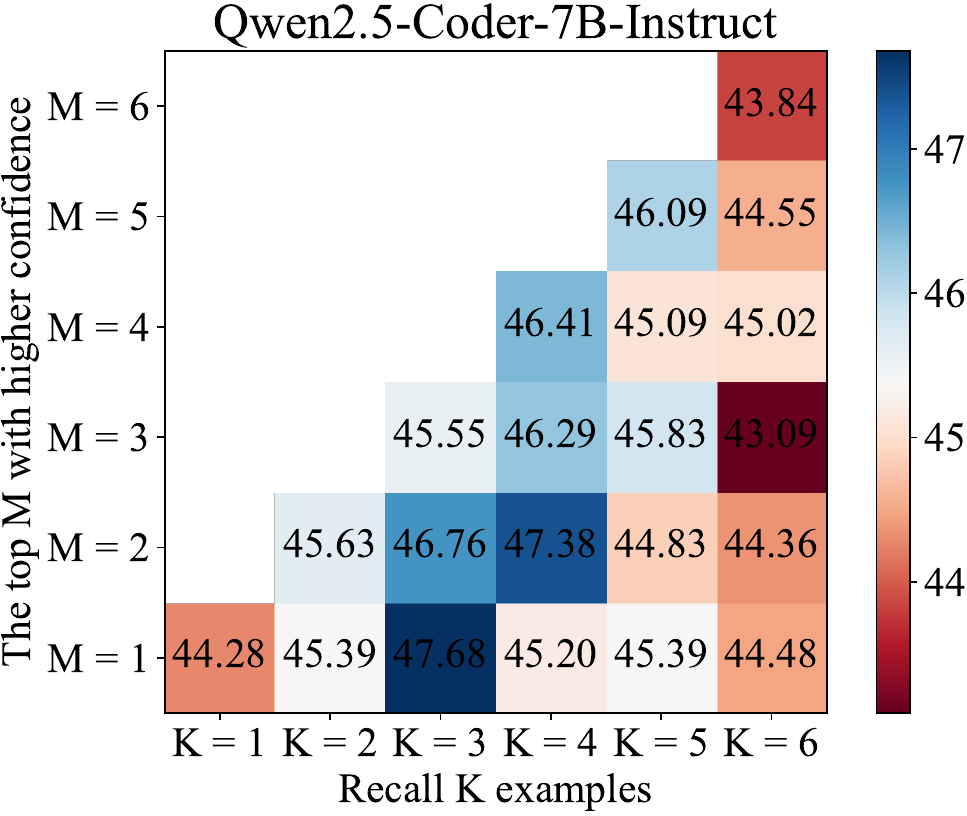}}
  \hspace{2.0mm}
 \subfloat[OOP]
  {\label{fig:stdy4}\includegraphics[scale=0.32]{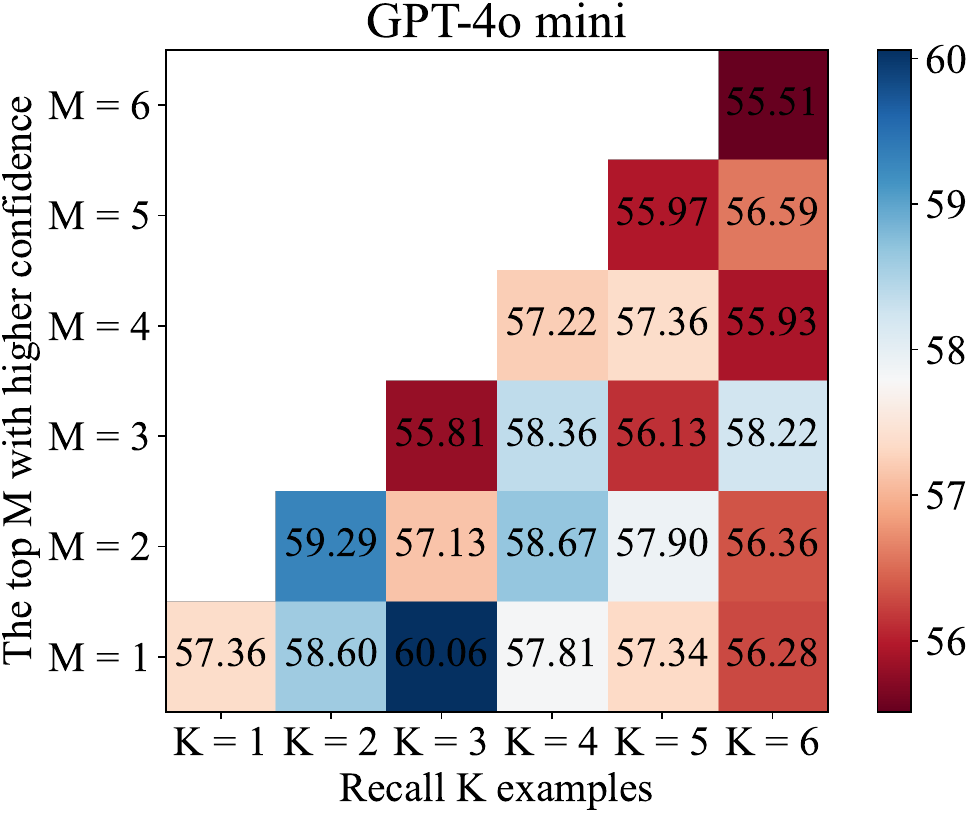}}
  \caption{The performance of LLMs (i.e., Qwen2.5-Coder-7B-Instruct and GPT-4o mini) based on the metamemory agent  on the HumanEval (top) and OOP (bottom) benchmarks when recalling $K$ examples and selecting the top $M$ recalled examples with the highest confidence.}
  \label{fig:study_result01}
\end{figure}

\begin{table}[htbp]
  \centering
  \caption{The accuracy of reference examples used by LLMs (i.e., Qwen2.5-Coder-7B-Instruct, InternLM2.5-7B-Chat and GPT-4o mini) based on analogical prompting and metamemory agent on the HumanEval and OOP benchmarks. Bold text indicates the best result. In the experiments, we used Ruff~\cite{wang2025refactoring,eghbali2025dylin} and Pylint~\cite{oliveira2022lint, eghbali2025dylin} as evaluation metrics.}
  \resizebox{0.99\linewidth}{!}{
    \begin{tabular}{cccccccc}
    \toprule
    \multirow{2}[4]{*}{LLMs} & \multirow{2}[4]{*}{Methods} & \multicolumn{2}{c}{HumanEval} & \multirow{2}[4]{*}{Avg} & \multicolumn{2}{c}{OOP} & \multirow{2}[4]{*}{Avg} \\
\cmidrule{3-4}\cmidrule{6-7}          &       & Ruff  & Pylint &       & Ruff  & Pylint &  \\
    \midrule
    \multirow{4}[2]{*}{Qwen2.5-Coder-7B-Instruct} & Analogical prompting & 76.39     & 29.63     & 53.01     & 92.78     & 47.61     & 70.20 \\
          & Metamemory agent ($K=3, M=3$) & 82.52     & 46.65     & 64.59     & 93.81     & 54.27     & 74.04 \\
          & Metamemory agent ($K=3, M=2$) & 88.65     & 55.61     & 72.13     & 95.82     & 65.71     & 80.77 \\
          & Metamemory agent ($K=3, M=1$) & \textbf{95.45}     & \textbf{67.37}     & \textbf{81.41}     & \textbf{97.43}     & \textbf{72.10}     & \textbf{84.77} \\
    \hdashline
    \multirow{4}[2]{*}{InternLM2.5-7B-Chat} & Analogical prompting & 88.27     & 42.65     & 65.46     & 92.41     & 52.93     & 72.67 \\
         & Metamemory agent ($K=3, M=3$) & 82.72     & 40.86     & 61.79     & 90.10     & 34.57     & 62.34 \\
          & Metamemory agent ($K=3, M=2$) & 87.45     & 50.32     & 68.89     & 93.27     & 49.68     & 71.48 \\
          & Metamemory agent ($K=3, M=1$) & \textbf{90.24}     & \textbf{58.71}     & \textbf{74.48}     & \textbf{95.11}     & \textbf{62.83}     & \textbf{78.97} \\
    \hdashline
    \multirow{4}[2]{*}{GPT-4o mini} & Analogical prompting & 87.60     & 34.58     & 61.72     & 92.91     & 53.26     & 73.09 \\
          & Metamemory agent ($K=3, M=3$) & 85.77     & 63.56     & 74.67     & 96.29     & 56.22     & 76.26 \\
          & Metamemory agent ($K=3, M=2$) & 90.04     & 70.04     & 80.04     & 98.05     & 65.24     & 81.65 \\
          & Metamemory agent ($K=3, M=1$) & \textbf{93.45}     & \textbf{75.34}     & \textbf{84.40}     & \textbf{99.23}     & \textbf{70.11}     & \textbf{84.67} \\
    \bottomrule
    \end{tabular}%
    }
  \label{tab:study_result02}%
\end{table}%

\subsection{Ablation Studies}
In this section, we assessed the role of the metamemory agent in enhancing reference example accuracy and investigated the effects of the number of recalled (i.e., $K$) and selected examples (i.e., $M$) on performance. The experimental results are shown in Table~\ref{tab:study_result02} and Figure~\ref{fig:study_result01}.

\subsubsection{RQ5. Authenticity and Accuracy of the Reference Example}
Taking Qwen2.5-Coder-7B-Instruct in Table~\ref{tab:study_result02} as an example, we can clearly see that the accuracy rates of the metamemory agent's reference examples on the HumanEval and OOP benchmarks are 81.41 and 84.77, respectively, which are 53.57\% and 20.75\% higher than those of analogical prompting. In InternLM2.5-7B-Chat and GPT-4o mini, the accuracy rates of the metamemory agent's reference examples also show significant improvements. Furthermore, when recalling three examples and selecting the one with the highest confidence, the metamemory agent can effectively eliminate or suppress recalled examples that may have logical flaws or semantic errors. For instance, in GPT-4o mini on the HumanEval and OOP benchmarks, when recalling three examples and choosing the one with the highest confidence, the accuracy rates of reference examples reach 84.40 and 84.67, respectively, which are 36.75\% and 15.84\% higher than those when using all recalled examples directly. These results indicate that the metamemory agent can significantly enhance the authenticity and accuracy of reference examples.

\noindent
\textbf{Answer to RQ5:} The proposed metamemory agent consistently ensures the authenticity and accuracy of reference examples in code generation tasks.

\subsubsection{RQ6. Impact of Recalled Examples}
In Figure~\ref{fig:study_result01}, we fix the value of $M$ to examine the impact of recalling $K$ relevant examples on performance. When $M=1$ and $3$, both Qwen2.5-Coder-7B-Instruct and GPT-4o mini exhibit a trend of first increase and then decrease performance on the HumanEval and OOP benchmarks. When $M=2$, their performance on the HumanEval benchmark steadily improves, while it declines on the OOP benchmark. These findings indicate that recalling more examples does not necessarily lead to the best result. Note: due to the input length limitations of the LLMs used in the experiments, we did not analyze the performance of recalling more examples.

\noindent
\textbf{Answer to RQ6:} Recalling more examples using a metamemory agent does not necessarily enhance the performance of LLMs on code generation tasks. On the contrary, it may even reduce their performance on one-time code generation tasks.

\subsubsection{RQ7. Impact of Selected Examples}
In Figure~\ref{fig:study_result01}, we fix the value of $K$ to examine the effect of selecting the $M$ recalled examples with higher confidence on performance. When $K=5$ and $6$, Qwen2.5-Coder-7B-Instruct exhibits a trend of first increasing and then decreasing performance on the HumanEval and OOP benchmarks. GPT-4o mini shows a similar trend on the HumanEval and OOP benchmarks. These findings indicate that selecting either more or fewer recalled examples does not necessarily improve the performance of the LLMs on one-time code generation tasks. Note: we did not analyze the cases of $K=2$, $3$, and $4$, as selecting fewer examples may not sufficiently validate the experimental findings.

\noindent
\textbf{Answer to RQ7:} In the metamemory agent, selecting either more or fewer recalled examples does not necessarily improve the LLM’s performance on one-time code generation tasks.

\begin{figure}[htbp]
  \centering
  \captionsetup[subfloat]{font=footnotesize,labelfont=rm,textfont=rm}
  \captionsetup[subfloat]{labelformat=empty}
  \subfloat[]{
    \label{fig:diss_title}
    \includegraphics[scale=1.8]{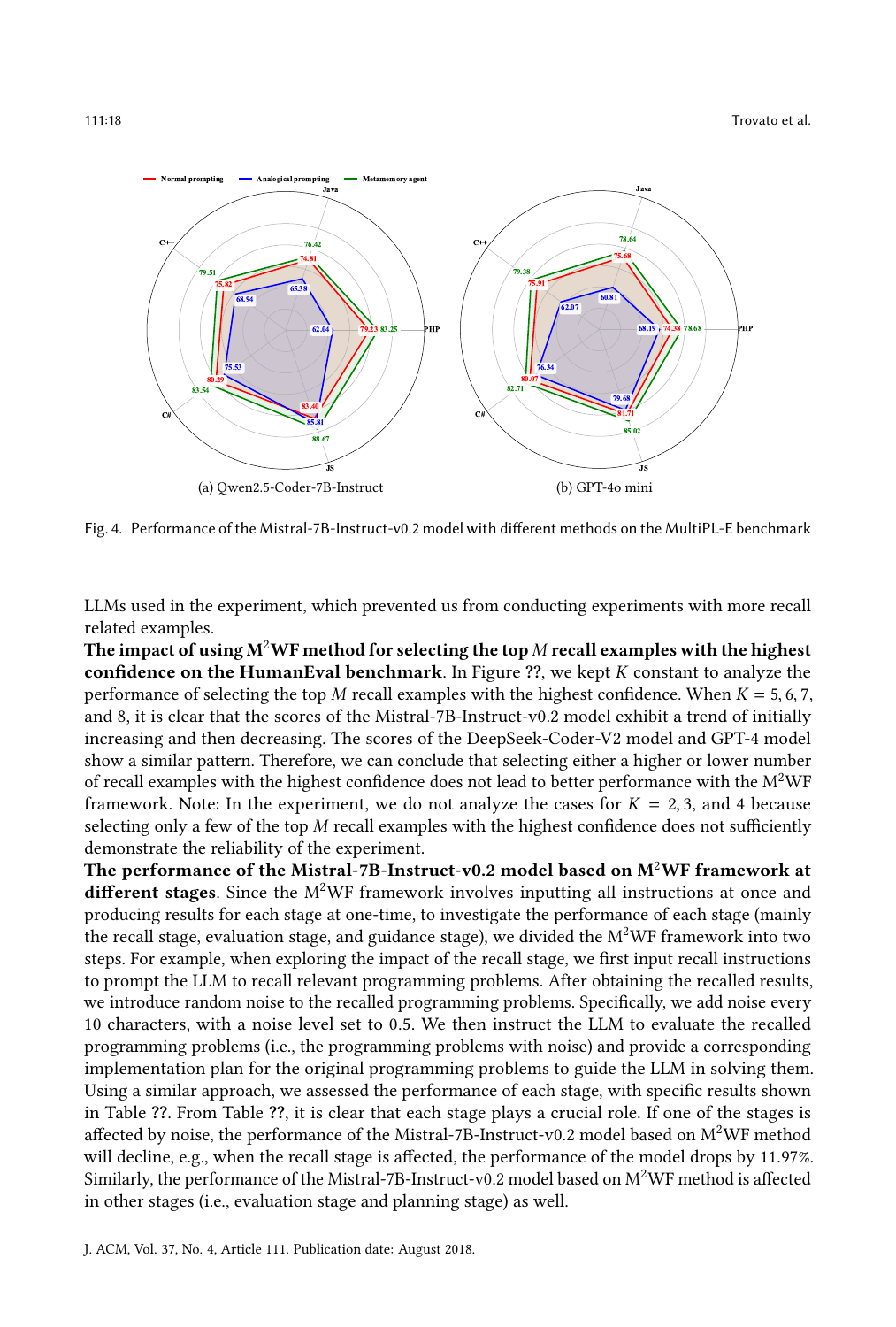}
  }
  \setcounter{subfigure}{0}
  \captionsetup[subfloat]{labelformat=parens}
  \\[-6mm]
  \subfloat[Qwen2.5-Coder-7B-Instruct]{
    \label{fig:diss_Qwen}
    \includegraphics[scale=0.30]{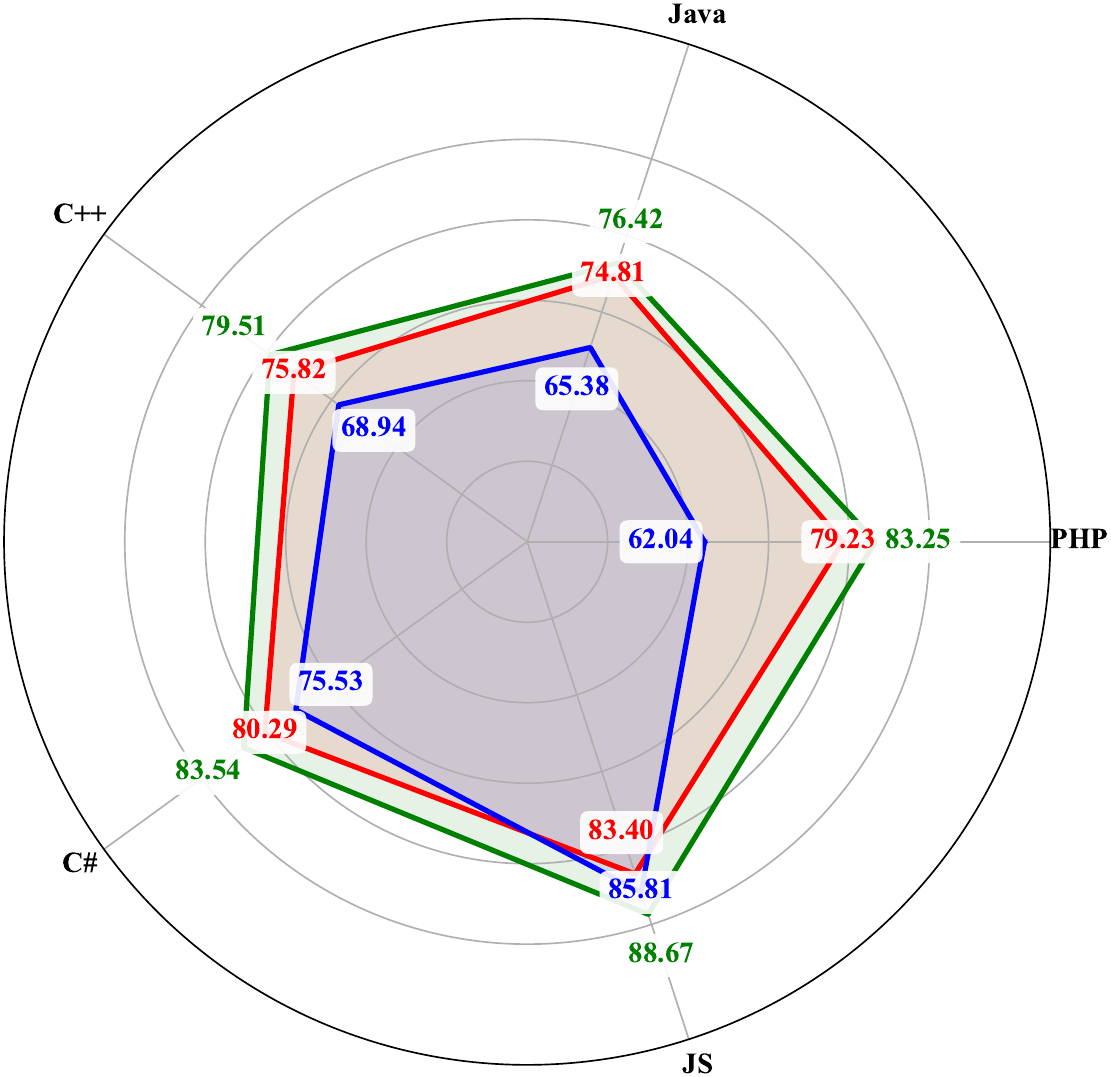}
  }
  \hspace{2.0mm}
  \subfloat[GPT-4o mini]{
    \label{fig:diss_GPT}
    \includegraphics[scale=0.30]{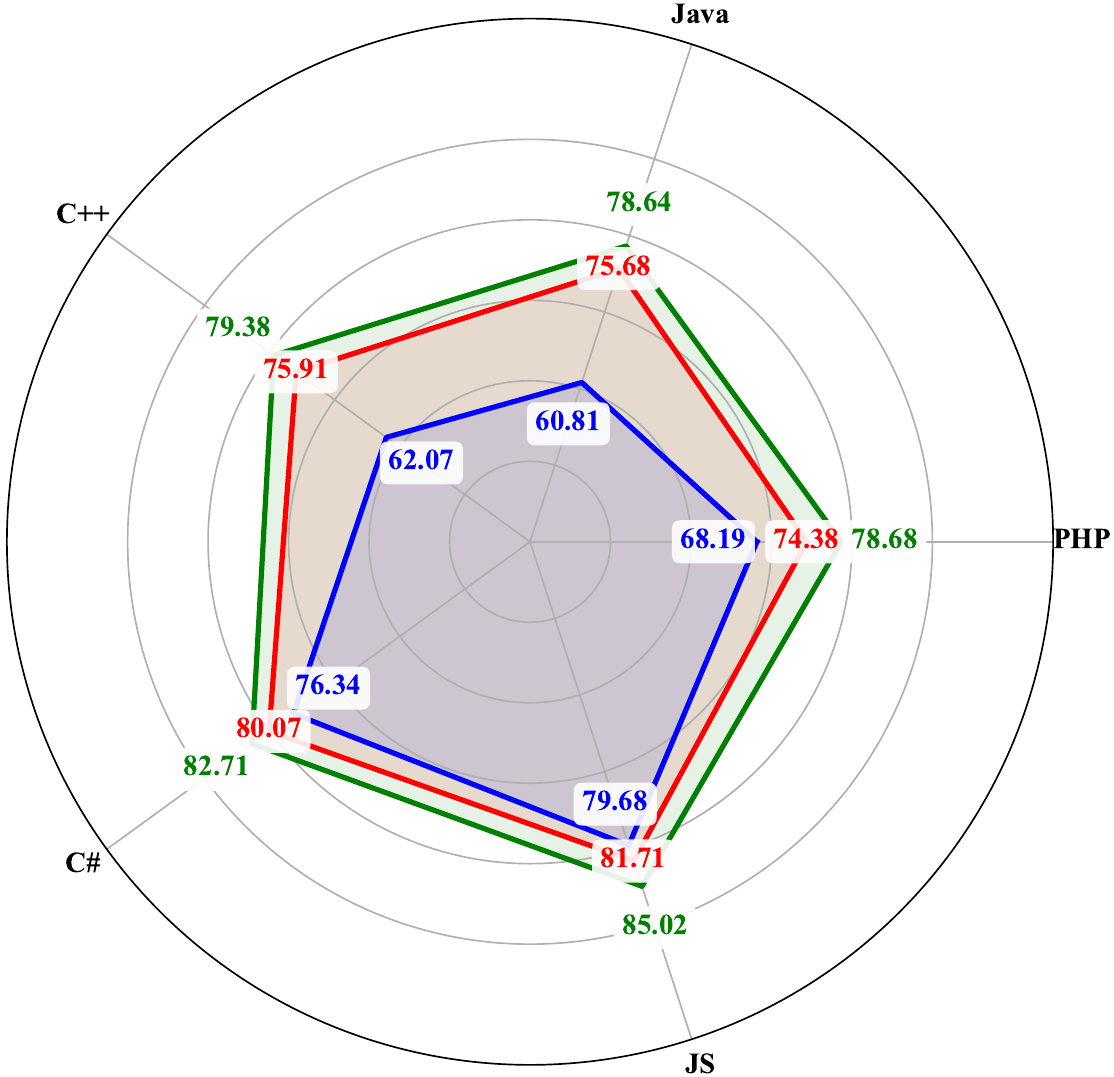}
  }
  \caption{Performance of the LLMs (i.e., Qwen2.5-Coder-7B-Instruct and GPT-4o mini) with different methods on the MultiPL-E benchmark. In the experiment, we set $K=3$ and $M=1$.}
  \label{fig:diss01}
\end{figure}

\begin{figure}[htbp]
    \centering
    \includegraphics[width=0.98\textwidth]{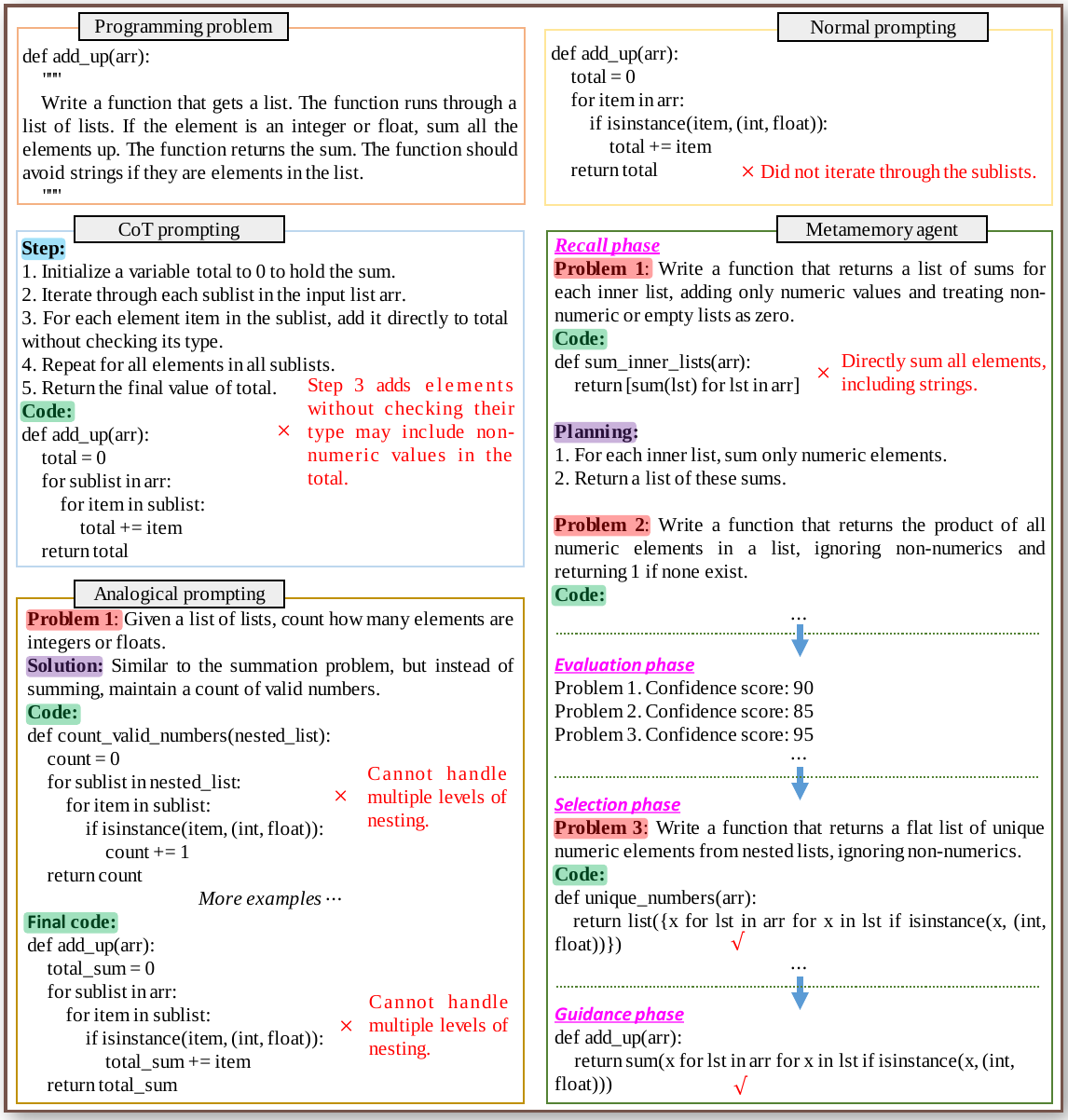}
    \caption{Examples of code generation using Normal prompting, CoT prompting, Analogical prompting, and Metamemory agent.}
    \label{fig:case_study}
\end{figure}

\subsection{Discussion}
In this part, we explore whether the metamemory agent can be applied to code generation tasks in other programming languages (i.e., C++, C\#, Java, JavaScript, and PHP) and analyze the reasons why it can improve the one-time code generation performance of LLMs. The experimental results are shown in Figures~\ref{fig:diss01} and~\ref{fig:case_study}.

\subsubsection{RQ8. Performance in Other programming languages}
As shown in Figure~\ref{fig:diss01}, the Qwen2.5-Coder-7B-Instruct with the metamemory agent achieves performance scores of 79.51, 83.54, 76.42, 88.67, and 83.25 on C++, C\#, Java, JavaScript, and PHP, respectively, representing improvements of 4.87\%, 4.05\%, 2.15\%, 6.32\%, and 5.07\% over Normal prompting. Similarly, GPT-4o mini with the metamemory agent surpasses Normal prompting by 4.57\%, 3.30\%, 3.91\%, 4.05\%, and 5.78\% across the same programming languages. These results indicate that the metamemory agent effectively enhances the one-time code generation performance of LLMs across multiple programming languages.

\noindent
\textbf{Answer to RQ8:} The metamemory agent can effectively improve the one-time code generation performance of LLMs in other programming languages.

\subsubsection{RQ9. Case study}
As shown in Figure~\ref{fig:case_study}, when generating code directly using Normal prompting, the LLM does not fully understand the specific requirements of the programming task, resulting in errors in the generated code, such as failing to iterate over sublists. When using CoT prompting, the LLM generates code step by step, but if an error occurs at any step, the final code will also be incorrect. For example, in Figure~\ref{fig:case_study}, during step 3 of CoT prompting, the LLM adds elements without checking their types, which may cause non-numeric values to be included in the total, leading to errors. Analogical prompting exhibits a similar issue: during the analogy process, the LLM may provide incorrect examples, which directly affect the correctness of the final code. As shown in Figure~\ref{fig:case_study}, the examples generated by analogical prompting cannot handle multi-level nesting, resulting in erroneous code. In contrast, our proposed metamemory agent can select accurate reference examples during the evaluation and selection process, enabling the LLM to generate correct code. Figure~\ref{fig:case_study} shows that even if incorrect examples are recalled, after evaluation and selection, the LLM ultimately uses the correct reference examples, ensuring that the generated code is error-free.

\noindent
\textbf{Answer to RQ9:} After recalling relevant examples, the metamemory agent can effectively eliminate recalled examples with logical flaws or semantic errors through the evaluation and selection process (i.e., ensuring the accuracy of the reference examples), enabling the LLM to generate correct code.

\begin{table}[htbp]
  \centering
  \caption{An overview of input and output tokens for ChatGPT and GPT-4o mini using different methods on the HumanEval and OOP benchmarks. The $\boldsymbol{\Delta}\left(\uparrow\right)$ indicates the relative token cost of CoT prompting, Plan-and-Solve prompting, Analogical prompting, and Metamemory agent with respect to normal prompting. Here, metamemory agent recall three related examples, i.e., a total of three evaluations were conducted.}
  \resizebox{1.0\linewidth}{!}{
    \begin{tabular}{ccccccccccc}
    \toprule
          & API Calls & Input & $\boldsymbol{\Delta}\left(\uparrow\right)$ & Output& $\boldsymbol{\Delta}\left(\uparrow\right)$ & API Calls & Input & $\boldsymbol{\Delta}\left(\uparrow\right)$ & Output & $\boldsymbol{\Delta}\left(\uparrow\right)$ \\
    \midrule   
    LLMs  & \multicolumn{5}{c}{ChatGPT}              & \multicolumn{5}{c}{GPT-4o mini} \\
    \midrule
    \multicolumn{11}{c}{HumanEval} \\
    \midrule
    Normal prompting & 1     & 23,344     & -     & 73,017     & -     & 1     & 23,344     & -     & 73,174     & - \\
    CoT prompting & 1     & 26,952     & 15.46\%     & 86,530     & 18.51\%     & 1     & 26,952     & 15.46\%     & 86,157     & 17.74\% \\
    Plan-and-Solve prompting & 1     & 30,396     & 30.21\%     & 104,136     & 42.62\%     & 1     & 30,396     & 30.21\%     & 104,807     & 43.23\% \\
    Analogical prompting & 1     & 48,786     & 108.98\%     & 158,800     & 117.48\%     & 1     & 48,786     & 108.98\%     & 158,240     & 116.25\% \\
    Metamemory agent & 6     & 526,655     & 2156.06\%     & 155,455     & 112.90\%     & 6     & 523,535     & 2142.70\%     & 153,649     & 109.98\% \\
    \midrule   
    \multicolumn{11}{c}{OOP} \\
    \midrule
    Normal prompting & 1     & 66,420     & -     & 35,933     & - & 1     & 66,420     & -     & 35,586     & -      \\
    CoT prompting & 1     & 75,905     & 14.28\%     & 43,234     & 20.32\% & 1     & 75,905     & 14.28\%     & 42,710     & 20.02\%      \\
    Plan-and-Solve prompting & 1     & 84,956     & 27.91\%     & 50,935     & 41.75\% & 1     & 84,956     & 27.91\%     & 50,589     & 42.16\%      \\
    Analogical prompting & 1     & 133,228     & 100.58\%     & 536,812     & 1393.92\%  & 1     & 133,228     & 100.58\%     & 538,999     & 1414.64\%      \\
    Metamemory agent & 6     & 1,479,291     & 2127.17\%     & 487,389     & 1256.38\%     & 6     & ,1479,953     & 2128.17\%     & 489,225    & 1274.77 \\
    \bottomrule
    \end{tabular}%
    }
  \label{tab:token_use}%
\end{table}%

\section{Threats to Validity}
\label{sec:threats_validity}
Although the metamemory agent demonstrates notable improvements in code generation tasks, several factors may threaten the validity of the results. First, when using the LLM’s API, the model may occasionally refuse to recall programming problems, potentially affecting task completion. Second, the metamemory agent comprises four phases (i.e., recall, evaluation, selection, and guidance), which substantially increases the number of input and output tokens (see Table~\ref{tab:token_use}). Third, the metamemory agent guides the LLM to simulate the human metamemory process through prompts, which makes the LLM highly sensitive to prompt phrasing and may limit the generalizability of the results.

\section{Conclusion}
\label{sec:conclusion}
Existing few-shot retrieval methods have achieved good results in one-time code generation tasks in LLMs. However, few-shot retrieval methods are particularly challenging for real-world data-free coding scenarios or benchmark tasks (e.g., HumanEval,
HumanEval+, OOP, StudentEval, and LiveCodeBench). In this work, inspired by human metamemory processes, we propose metamemory agent to improve the performance of one-time code generation in LLMs. This agent removes the dependency on retrieving examples from training data while maintaining the authenticity and accuracy of the LLM's recalled content. Extensive experiments demonstrate that the metamemory agent substantially boosts performance, and future work will explore its application in real-world software development to improve efficiency.



\section*{Funding}
This work is supported in part by the National Natural Science Foundation of China (Grant No. U23A20318) and the Science and Technology Major Project of Hubei Province (Grant No. 2024BAB046).

\section*{Declaration of competing interest}
The authors declare that they have no known competing financial interests or
personal relationships that could have appeared to influence the work reported
in this paper.

\printcredits

\section*{Data availability}
All datasets used in this study are publicly available benchmarks: HumanEval,
HumanEval+, MBPP, OOP, StudentEval, LiveCodeBench, Codeforces, and MultiPL-E,
each cited in the text. No new data were created in this study. The source code
will be made publicly available upon acceptance.

\section*{Declaration of Generative AI Use}
The author used generative artificial intelligence (AI) tools solely for language polishing during the preparation of this manuscript. The AI assistance was limited to improving grammar, wording, sentence fluency, and overall readability. No AI tools were used to generate research ideas, methodology, experimental design, data analysis, results, interpretations, or conclusions. All academic content, analyses, and final decisions were independently completed and verified by the author, who takes full responsibility for the accuracy and integrity of this manuscript.

\bibliographystyle{elsarticle-num}
\bibliography{sample-base}

\end{document}